\documentclass[society=none,twocolumn,superscriptaddress,amsmath,amssymb,floatfix]{revtex4-2}

\usepackage{times,bbm,amsmath,amssymb,amsthm}
\usepackage{braket}
\usepackage{epsfig,color}
\usepackage[colorlinks=true,citecolor=blue]{hyperref}
\hypersetup{
    allcolors  = {blue},
}
\usepackage{xcolor}
\usepackage{subfigure}
\usepackage{graphicx}
\usepackage[export]{adjustbox}
\usepackage{algorithm}
\usepackage{algpseudocode}
\usepackage{xpatch} 
\usepackage{xspace}
\makeatletter
\xpatchcmd\@collaboration@present{(}{\medskip}{}{}
\xpatchcmd\@collaboration@present{)}{}{}{}
\makeatother

\usepackage[titletoc,title]{appendix}
\usepackage{soul}

\begin{document}
\title{Interaction-free measurement of multiple objects using a universal integrated photonic processor}

\begin{abstract}
   The phenomenon of interaction-free measurement (IFM) enables the probabilistic detection of an absorbing object with reduced photon absorption. We report the experimental implementation of a generalized IFM of multiple objects using a single quantum probe on the cloud-based Ascella photonic processor developed by Quandela. We demonstrate sequential IFM of up to 5 emulated absorbers using a single photon, significantly extending the original IFM scheme for a single object. The experimental error-mitigated results confirm the theoretical predictions for this sequential IFM setup, and demonstrate a practical approach to scaling IFM to more complex quantum interrogation tasks.
\end{abstract}

\author{Sara Franco}
\thanks{Corresponding author. Email: \href{sara.franco@inl.int}{sara.franco@inl.int}}
\affiliation{International Iberian Nanotechnology Laboratory (INL), Av. Mestre José Veiga s/n, 4715-330 Braga, Portugal}
\affiliation{Centro de Física, Universidade do Minho, Braga 4710-057, Portugal}

\author{Anita Camillini}
\affiliation{International Iberian Nanotechnology Laboratory (INL), Av. Mestre José Veiga s/n, 4715-330 Braga, Portugal}
\affiliation{Centro de Física, Universidade do Minho, Braga 4710-057, Portugal}
\affiliation{CINECA Consorzio Interuniversitario, Via Magnanelli 6/3, 40033 Casalecchio di Reno, Italy}

\author{Ernesto F. Galvão}
\affiliation{International Iberian Nanotechnology Laboratory (INL), Av. Mestre José Veiga s/n, 4715-330 Braga, Portugal}
\affiliation{Instituto de Física, Universidade Federal Fluminense, Av. Gal. Milton Tavares de Souza s/n, Niterói, RJ, 24210-340, Brazil}

\maketitle

\makeatletter
\begingroup
\let\protect\@unexpandable@protect
\def\@xfootnotenext{*\kern-3pt}%
\protected@xdef\@thanks{\@thanks\protect\footnotetext
  [\the\c@footnote]{Corresponding author: \protect\href{mailto:email@domain.com}{email@domain.com}}}
\endgroup
\makeatother

\section{Introduction}

The phenomenon of ``negative-result" measurements was first introduced by Renninger in 1960, challenging the usual intuition about the nature of measurement in quantum mechanics. In his seminal work, Renninger showed that information about a quantum system's state can be obtained from the \textit{nonobservance} of a particular event, seemingly without any interaction occurring with the measured system \cite{renningerMessungenOhneStorung1960}. In 1981, Dicke discussed a related paradox: the \textit{nonscattering} of a photon off an atom can collapse the atom's wavefunction, an example of an ``Interaction-Free Measurement" (IFM) \cite{dickeInteractionfreeQuantumMeasurements1981}. This concept was firmly established in 1993 by Elitzur and Vaidman (EV), who proposed a now-famous interferometric “bomb-tester” thought experiment. Their protocol consists of a single particle, or probe (for instance, a photon) traversing a Mach-Zehnder interferometer where one path may be obstructed by an opaque object - a classical bomb, in their analogy. The interferometer is arranged for destructive interference at one output in the absence of the object. If the object is present, even the possibility of interaction disturbs the interference, causing the photon to end up at the “dark” output port with some probability. Detection of a photon at that port thus signals the object’s presence without any photon-object interaction \cite{elitzurQuantumMechanicalInteractionfree1993}. Shortly afterwards, Kwiat et al. introduced an improved IFM scheme \cite{kwiatInteractionFreeMeasurement1995} that exploits the quantum Zeno effect \cite{misraZenosParadoxQuantum1977,itanoQuantumZenoEffect1990} to in principle achieve arbitrarily high efficiency. In their 1995 proposal, a single photon weakly probes the presence of the object multiple times by undergoing several cycles through an interferometer. This suppresses absorption and boosts the probability of a successful interaction-free detection towards unity \cite{kwiatInteractionFreeMeasurement1995}. Subsequent experiments and analyses by Kwiat and collaborators tested and refined this high-efficiency approach \cite{kwiatExperimentalTheoreticalProgress1998,kwiatHighefficiencyQuantumInterrogation1999}.\par
These foundational works on IFMs - also known in the literature as ``quantum interrogation" protocols or ``counterfactual" measurements - have inspired several potential applications, such as counterfactual quantum key distribution schemes \cite{nohCounterfactualQuantumCryptography2009, liuExperimentalDemonstrationCounterfactual2012, salihProtocolDirectCounterfactual2013}, interaction-free imaging techniques \cite{whiteInteractionFreeImaging1998,lemosQuantumImagingUndetected2014,yangInteractionfreeSinglepixelQuantum2023,zhangInteractionfreeGhostimagingStructured2019,hanceCounterfactualGhostImaging2021,kentQuantumInterrogationSafer2001} or exchange-free or counterfactual quantum gates and computation paradigms \cite{salihExchangefreeComputationUnknown2021,salihDeterministicTeleportationUniversal2025}. These applications show IFM is not only a fundamental quantum-mechanical effect, but also a useful quantum resource for obtaining information with minimal disturbance.\par
To date, most experimental realizations of IFMs have employed single-photon probes and free-space optical setups \cite{kwiatInteractionFreeMeasurement1995, kwiatExperimentalTheoreticalProgress1998, kwiatHighefficiencyQuantumInterrogation1999, whiteInteractionFreeImaging1998,lemosQuantumImagingUndetected2014,yangInteractionfreeSinglepixelQuantum2023,caoCounterfactualUniversalQuantum2020}. Implementing these interferometric schemes in bulk optics presents significant challenges. High-visibility interference fringes require active phase stabilization on sub-wavelength scales, especially for the multi-pass high-efficiency setups, and optical components inevitably introduce loss and alignment errors. Integrated photonic platforms offer a promising alternative, as monolithic integration provides inherent phase stability, miniaturization, and scalability not achievable in large-scale optical tables \cite{metcalfMultiphotonQuantumInterference2013,wangIntegratedPhotonicQuantum2020}. Indeed, a photonic on-chip implementation of a high-efficiency IFM was demonstrated in 2014 by Ma et al. \cite{maOnchipInteractionfreeMeasurements2014}, achieving excellent spatial mode matching and stable interferometric phase. More recently, Giordani et al. implemented the standard EV interrogation task on a programmable Universal Photonic Processor (UPP) \cite{giordaniExperimentalCertificationContextuality2023}. This device benefits from the stability of integration while also offering tunability and rapid reconfiguration, enabling quantum protocols like IFM to be realized without manual realignment. The technological advancement of these devices has made possible an interesting new kind of tool - cloud-based services which make fully-functioning UPPs available remotely to the scientific community. Such tools greatly facilitate experimental exploration and proof-of-principle tests of photonic protocols and algorithms. Taking advantage of this possibility, in this work we report the experimental realization of the standard single-object EV IFM using the Ascella photonic quantum processor, a remotely-accessible UPP made available via the Quandela Cloud service. We benchmark the performance of the device using this well-known task, and discuss experimental error-mitigation techniques relevant for this type of platform.  \par
An as yet largely unexplored avenue of IFMs is the possibility of detecting multiple objects with a single quantum probe. In 2024, Filatov and Auzinsh \cite{filatovSetupInteractionfreeMeasurement2024} put forward a proposal which takes advantage of the fact that, since the probe particle is not destroyed during an interaction-free detection, it can be reused to interrogate additional objects sequentially, in an overlapping scheme of IFM devices capable of detecting the presence of several objects with a single detection of the probe. Motivated by this idea, we demonstrate an experimental realization of an interaction-free measurement involving multiple objects. Exploiting the universal architecture of Ascella UPP, suitable for the implementation of arbitrary linear optical circuits on up to 12 modes, we implement an adaptation of the Filatov-Auzinsh proposal suitable for an on-chip implementation, which is capable of recycling a single photon for successive measurements, generalizing the IFM protocol to multiple absorbers. This work goes beyond the theoretical proposal of Filatov and Auzinsh by achieving a practical demonstration of multi-object IFM on a functional photonic quantum processor.\par
This paper is organized as follows: Sec.\ref{section:background} provides a theoretical background on IFMs, from the original EV proposal, Kwiat et al.'s high-efficiency scheme and their proposed practical applications to generalizations of these tasks to setups with higher number of modes and objects, including, in particular, the Filatov-Auzinsh proposal. In Sec.\ref{section:our_multiple_object}, we introduce a non-overlapping scheme for quantum interrogation of multiple objects, which in contrast to the Filatov-Auzinsh proposal, is suitable for an on-chip architecture with no spatial overlap between optical modes and with no control over other degrees of freedom of photons. In Sec.\ref{section:exp} we report on our experimental implementations of the standard EV task, as well as our multiple object generalization, on the cloud-accessible Ascella photonic processor. We conclude with a discussion of our results in Sec.\ref{section:discussion}.

\section{Background}\label{section:background}

In this section, we provide some background on IFM protocols. Sec.\ref{section:EV} reviews the original EV proposal for an IFM of a single object, also referred to in this work as the standard quantum interrogation task, and introduces the protocol's efficiency as a figure of merit. Sec. \ref{section:multimode} describes generalized single-object IFM tasks inspired on the EV proposal, with efficiency improvements or added functionalities. In Sec.\ref{section:Filatov}, we review a recent proposal that extends the concept of IFM to the quantum interrogation of multiple objects using a single probe.

\subsection{Standard quantum interrogation}\label{section:EV}

\begin{figure}[t]
    \centering
    \includegraphics[]{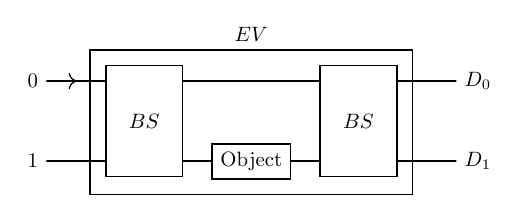}
    \caption{The Elitzur-Vaidman (\textit{EV}) setup for an interaction-free measurement, based on a Mach-Zehnder interferometer. Single photons are input on the left in mode 0, and exit towards photodetectors $D_0$ or $D_1$. The beamsplitters BS of reflectivity $R$ act on the incoming optical modes according to eq.\eqref{eq:BS}. The interferometer is configured for destructive interference in output mode 1 in the absence of obstruction. When an opaque object is placed in one of the arms, for example, the bottom one, detection at $D_1$ signals the presence of the object in the device, without an interaction occurring between it and the photon.}
    \label{fig:EV}
\end{figure}

The standard EV protocol for an IFM \cite{elitzurQuantumMechanicalInteractionfree1993} is based on a Mach-Zehnder Interferometer (MZI), depicted in Fig.\ref{fig:EV}. Each Beam Splitter (BS) enacts the following transformation on the creation operators associated with input modes
\begin{equation}\label{eq:BS}
    BS(R) = 
    \begin{pmatrix}
        \sqrt{R} & \sqrt{T}\\
        \sqrt{T} & -\sqrt{R}
    \end{pmatrix},
\end{equation}
\noindent where $R$ is the BS reflectivity, and $T = 1-R$. With this definition, a reflected photon will exit through the same mode it entered the BS, while a transmitted photon will exit through the other mode. Single photons are input in port 0, and photodetectors $D_0$ and $D_1$ record photon counts at each output port. The interferometer is configured such that, in the absence of the object, destructive interference occurs at output port 1 (also called the dark port) and only detector $D_0$ clicks. The presence of an opaque classical object in one of the arms of the MZI, for instance, the bottom one, inhibits interference, changing the possible measurement outcomes. After the first BS, there is a probability $P_{\rm abs} = T$ that an incoming photon takes the lower path, in which case it is absorbed by the object. With a probability $P_{light} = R^2$, the photon avoids the object and ends up at the top, or light, port. This outcome provides no information on the presence of the object. Finally, there is a non-zero probability $P_{\rm IFM} = RT$ that the photon ends up at the dark port. It follows that $D_1$ only clicks if the object is present in the arm of the device. If the photon reaches the detector, it cannot have been absorbed by the object. Therefore, a click at $D_1$ corresponds to an IFM of its presence in the MZI arm. \par
Several experimental implementations of the EV protocol have been realized, some with an equivalent, Michelson interferometer version of the original MZI design, using single photons \cite{kwiatInteractionFreeMeasurement1995, giordaniExperimentalCertificationContextuality2023}, classical light beams \cite{dumarchievanvoorthuysenRealizationInteractionfreeMeasurement1996} and neutron interferometry \cite{hafnerExperimentInteractionfreeMeasurement1997}. In \cite{kwiatExperimentalTheoreticalProgress1998,whiteInteractionFreeImaging1998}, the EV scheme is used to perform optical quantum imaging of several classical objects, such as metallic wires and knife edges. By scanning the object through the interferometer arm and monitoring the dark port, it is possible to reconstruct one-dimensional profiles of the object, where the imaging photons are precisely the ones that \textit{do not} scatter off of the object. More sophisticated interaction-free imaging setups inspired on the EV scheme have since been proposed \cite{lemosQuantumImagingUndetected2014,yangInteractionfreeSinglepixelQuantum2023}, including some that consider the imaging of semi-transparent objects \cite{kentQuantumInterrogationSafer2001,paliciInteractionfreeImagingMultipixel2022}. EV's proposal has also inspired quantum key distribution schemes of enhanced security, where information-carrying particles are not exchanged between the parties \cite{guoQuantumCryptographyBased1999,nohCounterfactualQuantumCryptography2009,liuExperimentalDemonstrationCounterfactual2012}.\par
In the EV protocol, probes are sent one at a time through the MZI in succession, until either an absorption by the object or an IFM detection occurs. Typically, the task is considered successful only when a click at the dark port occurs before an object absorption. A common figure of merit used to characterize the performance of the protocol is the efficiency
\begin{equation}\label{eq:eta_EV}
    \eta  = \frac{P_{\rm IFM}}{P_{\rm IFM} + P_{\rm abs}},
\end{equation}

\noindent defined as the probability that a successful IFM is achieved assuming an infinite amount of trials can be attempted. In terms of the BS reflectivity, it reads
\begin{equation}\label{eq:eta_EV_R}
    \eta(R) = \frac{R}{R+1},
\end{equation}

\noindent or, equivalently,
\begin{equation}\label{eq:eta_EV_T}
    \eta(T) = \frac{1-T}{2-T},
\end{equation}

\noindent an increasing function of $R$, asymptotically approaching an upper bound of $0.5$ as $R \to 1$. There is, however, some subtlety when dealing with this limit. When $R = 1$, the BSs become perfect mirrors, and the photon no longer has a chance of interacting with the object, rendering an IFM impossible. This reflects the fact that $P_{\rm IFM}$, the absolute probability of an interaction-free detection per input photon, is maximized for $R=0.5$ and decays to zero as $R \to 1$. Consequently, as we approach this limit, an increasing number of single photons need to be sent on average through the MZI before an IFM occurs, with most photons ending up at the light port. In other words, the efficiency can be large even if $P_{\text{IFM}}$ is close to zero, as long as $P_{\text{abs}}$ is also negligible, since $\eta$ does not quantify how many attempts are required to detect the object, but only the probability that \textit{eventually} it will be found, interaction-free. When considering potential applications of this protocol, other figures of merit might be more relevant. For example, one might be interested in maximizing the success probability given a maximum number of allowed trials $k$, in which case the optimal reflectivity actually decreases as we decrease $k$. Alternatively, in an imaging context, one might be interested in boosting $P_{\text{IFM}}$ to speed-up the imaging process at the cost of allowing more absorptions to occur, in which case $R=0.5$ might be the optimal choice. Finally, considerations about the sensitivity of $\eta$ to experimental errors might also impact the optimal choice for the reflectivity $R$. While efficiency does not fully capture practical feasibility, we adopt it here because it is standard in the literature and provides a hardware- and application-agnostic measure of the protocol's performance.\par

\subsection{Generalized single object quantum interrogation}\label{section:generalized_single_object}

This section introduces later developments on single-object IFM protocols, inspired by the original EV proposal. Sec.\ref{section:kwiat} briefly describes a scheme that reaches arbitrarily high efficiency by encoding information on the photon's polarization via the Quantum Zeno effect. Sec.\ref{section:multimode} discusses protocols which make use of a higher number of spatial modes.

\subsubsection{High-efficiency quantum interrogation}\label{section:kwiat}

In \cite{kwiatInteractionFreeMeasurement1995,kwiatExperimentalTheoreticalProgress1998,kwiatHighefficiencyQuantumInterrogation1999}, Kwiat et al. propose an IFM protocol that breaks the $50\%$ efficiency ceiling of the EV setup, by combining it with the discrete quantum Zeno effect (QZE) \cite{misraZenosParadoxQuantum1977,itanoQuantumZenoEffect1990}. In their protocol, the BSs of Fig.\ref{fig:EV} are replaced with polarizing BSs (PBSs) that reflect horizontal polarization, and a polarization rotator is placed in mode 0 just before the input. A horizontally polarized single photon is sent through the rotator, which rotates its polarization by $\pi/2N$, with $N$ a positive integer, and then passes through the MZI with PBSs; instead of heading towards photodetectors, the photon is rerouted to the input, and repeats this cycle $N$ times in total, interrogating the object $N$ times. If the object is absent, the photon's polarization deterministically evolves to vertical after the $N$ cycles. If the object is present, the photon has a probability $P=\sin^2(\pi/2N)$ of being absorbed at every round, and survives until the last cycle with probability $P_{\text{IFM}}=\cos^{2N}(\pi/2N)$. At every interrogation, the non-absorption by the object measures the photon's polarization to be horizontal, ``freezing" the evolution of the polarization state via the QZE. By measuring the photon's polarization at the end, definite information is thus obtained on the presence, or absence, of the object. As we increase $N$, $P_{\text{abs}}$ vanishes and the probability $P_{\text{IFM}}$ that the photon survives the $N$ cycles approaches unity, and therefore, the efficiency $\eta = \cos^{2N}(\pi/2N)$ can be made arbitrarily close to unity.\par
By reaching, in principle, arbitrarily high efficiency, Kwiat et al.'s proposal is even more promising for practical applications than the EV setup, albeit also proving more challenging to implement due to its added complexity. This higher efficiency version has inspired applications to quantum imaging \cite{hanceCounterfactualGhostImaging2021} and communications \cite{salihProtocolDirectCounterfactual2013} and, if a quantum object is used instead of a classical one (a possibility already explored in EV's original paper \cite{elitzurQuantumMechanicalInteractionfree1993}), it could in principle be used to implement exchange-free 2-qubit quantum gates, with potential use in distributed quantum computing tasks \cite{salihDeterministicTeleportationUniversal2025,salihExchangefreeComputationUnknown2021}.
 
\subsubsection{Multimode setups}\label{section:multimode}

\begin{figure}
    \centering
    \includegraphics[]{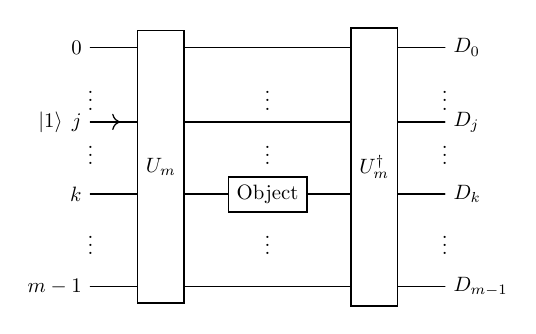}
    \caption{Generalized setup for single object interaction-free measurement. Any $m$-mode interferometer implementing a generic unitary transformation $U_m$ followed by its inverse $U_m^{\dagger}$ can be used to perform the interrogation of a single object, placed in any mode $k$ between the two unitaries. If single photons are input in mode $j$, they always exit towards photodetector $D_j$ at the output when the object is absent. A detection at any output mode but $j$ heralds the presence of an object in an undetermined arm of the interferometer.}
    \label{fig:UUdagger}
\end{figure}

A natural generalization of the EV protocol to be considered is the extension to setups with a higher number of spatial modes. In fact, quantum interrogation of a single object can be realized with any linear $m$-mode interferometer implementing a generic unitary transformation $U_m$ followed by its inverse $U_m^{\dagger}$. Consider the scheme depicted in Fig.\ref{fig:UUdagger}, where a single photon is sent through input mode $j$ of an $m$-mode interferometer. In the absence of any object, the net effect $U_m U_m^\dagger$ is the identity, and the photon will exit through port $j$. However, if an opaque object is present in one of the inner arms (say, arm $k$ of $m$), the photon may be absorbed - if it propagates through arm $k$ - or it may be deflected into a different output mode - if the interference is disturbed by the potential absorption. Thus, a click at any detector but $D_j$ consists of an IFM of the presence of the object in some (undetermined) arm. The EV scheme in Fig.\ref{fig:EV} can therefore be seen as a special case of this most general construction, with $m=2$ and $U_2 = BS(R)$.\par
We propose an extension of the definition of the efficiency $\eta_{U_m}$ for this type of setup in terms of eq.\eqref{eq:eta_EV}, where $P_{\text{IFM}}$ is the probability of a click at any detector but $D_j$, and $P_{\text{abs}}$ the absorption probability. It turns out that $\eta_{U_m}$ depends only on the modulus square of the matrix element $[U_m]_{k,j}$ that couples the input mode $j$ of the photon to the mode $k$ where the object is placed,

\begin{equation}\label{eq:eta_U}
    \eta_{U_m} = \frac{1 - |[U_m]_{k,j}|^2}{2- |[U_m]_{k,j}|^2}.
\end{equation}

It is straightforward to check that this formula reduces to eq.\eqref{eq:eta_EV} when $|[U_m]_{k,j}|^2 = |[BS(R)]_{1,0}|^2 = T$. Comparing eqs. \eqref{eq:eta_EV_T} and \eqref{eq:eta_U}, we can therefore conclude that while the schemes in Figs. \ref{fig:EV} and \ref{fig:UUdagger} represent distinct physical implementations of an IFM, requiring different optical components and different numbers of optical modes, formally, their efficiency depends in the same manner on a single, tunable parameter, corresponding to the coupling transmission between the input mode of the probe and the mode obstructed by the object. There is, in principle, no gain in efficiency from implementing a multimode scheme, since all possible values of $\eta$ reachable in a higher mode scheme can likewise be spanned by tuning the reflectivity $R$ in the 2-mode EV protocol.\par
The higher the number of modes used, the less precise the information on the location of the object, since an IFM outcome does not allow us to determine which mode the object is obstructing. This indetermination would, for example, result in poorer resolution of the obtained image in a quantum imaging scenario. Thus, at first sight, there seems to be no advantage in implementing this multimode approach. However, there could be other practical advantages to be gained from using a higher mode scheme. For instance, in App.\ref{appendix:robustness}, we discuss a potential advantage of multimode schemes in experimental setups where each optical component is subject to independent errors. We present simulated data suggesting that multimode schemes might be less sensitive to such errors when approaching the efficiency ceiling, leading to reduced noise in that regime.\par
One way to determine the position of the object in a multimode structure is with a related protocol proposed in \cite{paliciInteractionfreeImagingMultipixel2022}, albeit with some key differences from the one described above. In this protocol, a photon in spatial mode 0 is prepared in a superposition of $m$ different orbital angular momentum (OAM) states, and instead of a unitary $U_m$ acting on the spatial modes, a unitary path-to-OAM encoder $S_m$ demultiplexes the OAM modes into $m$ different paths. After interrogating the object, which can be in any of the $m$ arms, analogously to the setup in Fig.\ref{fig:UUdagger}, the OAM components are multiplexed back to the same path 0 by an inverse $S^{\dagger}_m$. This setup is then embedded into an EV-type scheme, where instead of being obstructed by the single classical object, the lower arm of the MZI is connected to the input and output of mode 0 of this structure. At the output of the MZI, a demultiplexer is added to the dark port, with a photodetector for each of the $m$ demultiplexed modes. In this way, a click at any of the demultiplexed dark ports signals the detection of the object, with the OAM degree of freedom carrying the information on which arm the object is located. As the authors propose, such a scheme could be used to image the structure of a multi-pixel object with no scanning required. Each probe sent through the device can, with some probability, interrogate one pixel of the object. After enough collection events, a full image of the object can thus be reconstructed with a static setup. As the authors point out, by combining this idea with Kwiat et al.'s scheme described in Sec.\ref{section:kwiat}, a more efficient protocol can be obtained. While the photonic platform we consider in this work does not enable the manipulation of other degrees of freedom of photons besides their spatial mode, bulk optics and other setups could enable the experimental implementation of this type of system.\par
At this stage, we would like to highlight an important distinction between the proposal just described for imaging multiple pixels and the multi-object setups described in the upcoming sections. In the former, each probe is only capable of interrogating a single pixel of the object, with a certain probability. If the object contains 3 pixels, 3 different detection events, corresponding to a click in each of the 3 dark port detectors, need to occur before the object's structure is fully characterized. This scheme can thus be understood as $m$ EV experiments conducted in parallel, which is why we opt to classify it here as a single-object IFM. In contrast, the schemes we describe in Secs.\ref{section:Filatov} and \ref{section:our_multiple_object} are capable of detecting multiple pixels, or multiple objects, \textit{with a single probe}. A single detection event, occurring with a certain success probability, heralds the presence of all objects simultaneously.\par

\subsection{Quantum interrogation of multiple objects}\label{section:Filatov}

\begin{figure}
    \centering
    \includegraphics[width=\linewidth]{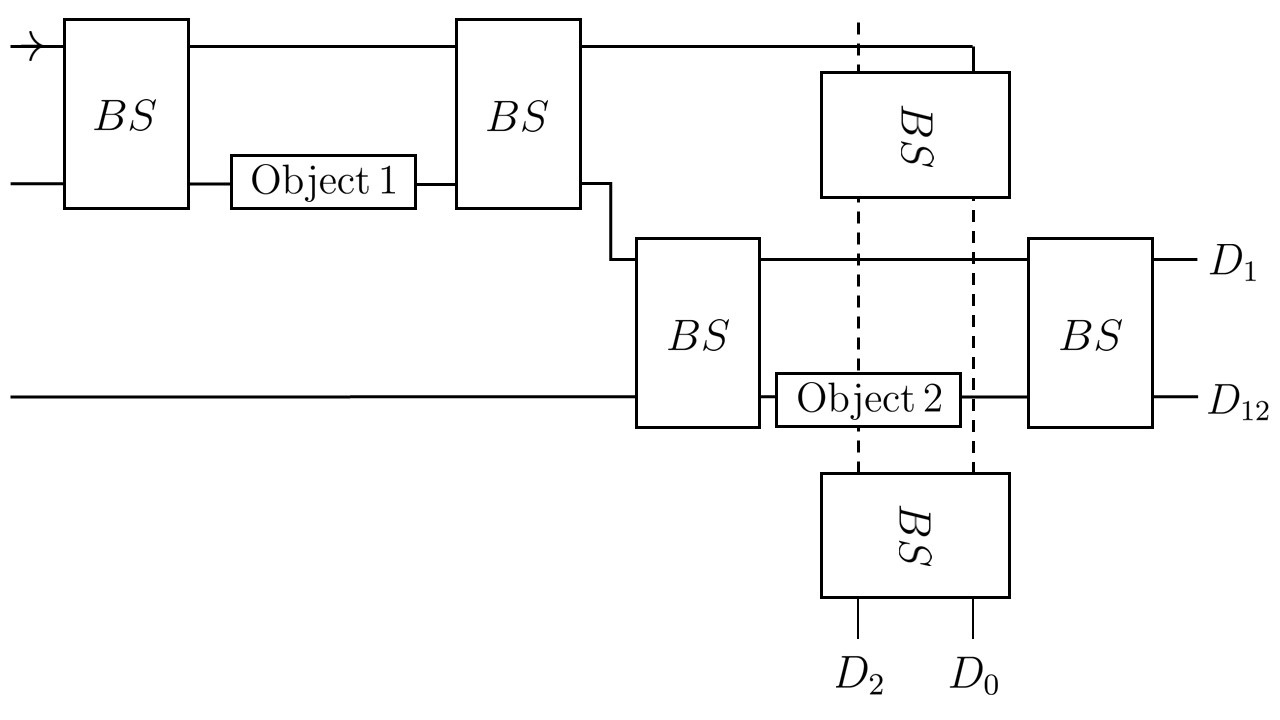}
    \caption{Filatov-Auzinsh proposal for the IFM of two objects using a single probe. The outputs of the EV scheme of Fig.\ref{fig:EV}, used for the IFM of object 1, are each redirected to one of two additional EV units, with spatially overlapped optical modes so that they can both interrogate object 2. Single photons are input at the top port of the leftmost EV unit. If detection occurs at photodetector $D_{12}$, the presence of both objects is simultaneously signalled with no interaction. Detections at $D_1$ or $D_2$ give partial information on the presence of object 1 or object 2, respectively. $D_0$ is the null outcome, giving no information on either object.}
    \label{fig:Filatov}
\end{figure}

In 2024, Filatov and Auzinsh first proposed a conceptual scheme to detect multiple objects interaction-free using a single probe particle \cite{filatovSetupInteractionfreeMeasurement2024}. Their proposal takes advantage of the fact that the probe - a photon - is preserved during interrogation of the object's presence, and can thus be used for subsequent interrogations. In a conceptual design for two objects, illustrated in Fig.\ref{fig:Filatov}, each output of an EV interferometer, possibly containing object 1, connects to one of two other EV setups. The latter two are partially overlapped, in such a way that both can potentially be obstructed by object 2. The photon initially goes through the first interferometer to test for object 1. Instead of terminating its journey at dark-port or bright-port detectors, the photon continues towards one of the two subsequent interferometers, probing object 2. In total, the system has four output ports where photodetectors are placed, yielding four possible detection outcomes: a click at $D_{12}$, indicating “object 1 and object 2 are both present”; at $D_1$, indicating “object 1 is present but no information about object 2”; at $D_2$, indicating “object 2 is present but no information about object 1”; and at $D_0$, giving no information about either. In effect, a single detection event of the probe at $D_{12}$ heralds the presence of both objects at their respective locations.\par
For the generalisation to $n$ objects, one could envision $2^n$ different outcomes corresponding to all combinations of presence/absence information – though implementing such a fully overlapping interferometer network would require an exponential number of physical resources if done spatially, as the $n^{th}$ additional object would require $2^{n-1}$ additional overlapping devices. Filatov and Auzinsh pointed out that, by encoding the information about each object in the photon’s temporal degree of freedom - e.g., by introducing a delay in the path at the exit of the dark port with respect to the bright one  - one could achieve the multi-object IFM with only linear resource scaling. In this case, the number of objects that can be detected is limited by the detector's time resolution and coherence length of the photons, and more effort is required in accounting for the photons' time-of-arrival in the results \cite{filatovSetupInteractionfreeMeasurement2024}.

Their proposal is probabilistic – the photon might still be absorbed by any of the objects – but if a detection event occurs in the designated output, it confirms the presence of all $n$ objects with no interactions, which is a striking generalization of the single-object EV effect. Furthermore, as the authors pointed out, if the EV units are replaced with the high-efficiency IFM device of Kwiat et al. (see Sec.\ref{section:kwiat}), 
near-definite information is obtained on the objects no matter the outcome, and with a significantly improved efficiency: essentially, in the limit of ideal operation, any detection of the photon gives definite information on the presence or absence of each object, with absorption events becoming exceedingly rare. The trade-off is that one must incorporate many repeated cycles for each object - as per Kwiat’s scheme - making the setup more complex. Filatov and Auzinsh’s work, however, remained a theoretical blueprint, and no experiment demonstrating even the basic multi-object IFM has, to our knowledge, been reported prior to our work.

\section{Non-overlapping scheme for multiple-object quantum interrogation} \label{section:our_multiple_object}

\begin{figure*}[t]
    \centering
     \subfigure[]{\includegraphics[width=0.45\textwidth,valign = m]{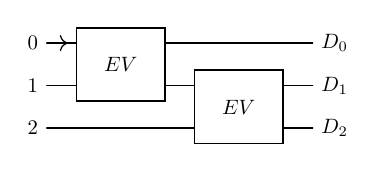}\label{fig:two_objects}}
    \hspace{0.5cm}
    \subfigure[]{\includegraphics[width=0.5\textwidth,valign = m]{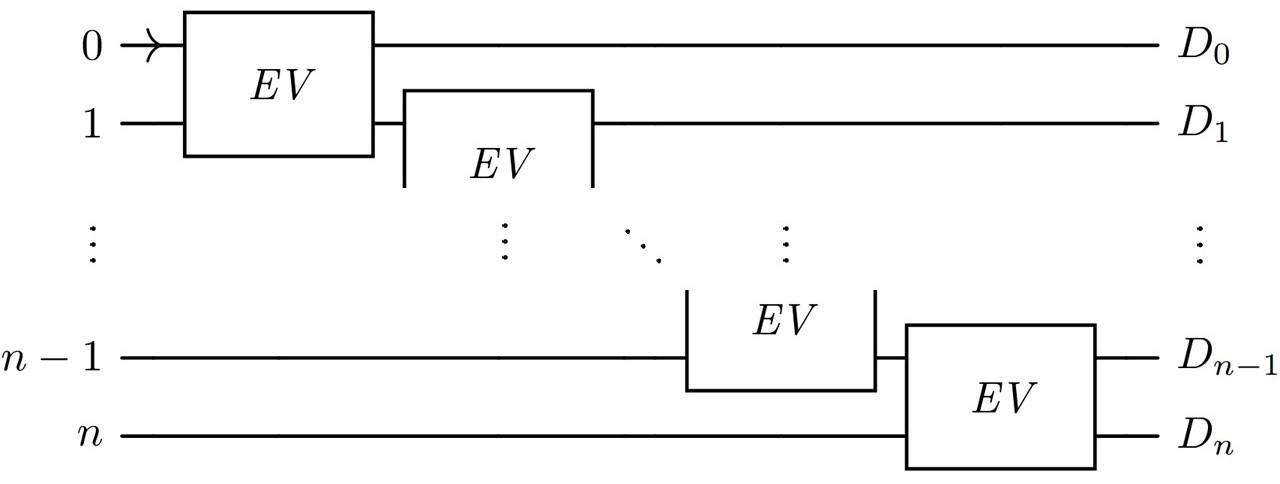}\label{fig:n_objects}}
    \caption{(a) Scheme for the interaction-free measurement of two objects using a single probe. Each EV box corresponds to the scheme in Fig.\ref{fig:EV}. Detection at $D_2$ only occurs when there is one object inside each box. This setup can be straightforwardly generalized to any number of objects. (b) Generalized scheme for the interaction-free measurement of $n$ objects using a single probe. Each EV box corresponds to the setup in Fig.\ref{fig:EV}, with some choice of the beamsplitters reflectivity $R$.}
\end{figure*}

The multi-object IFM scheme of Filatov-Auzinsh detailed in Sec.\ref{section:Filatov} requires an exponentially increasing number of spatially overlapping interferometers, or otherwise, the manipulation of at least two degrees of freedom of photons (for instance, spatial mode and temporal degree of freedom). Inspired by the principle of recycling the probe in further measurements, in this section we propose a simpler protocol that still captures the essence of Filatov and Auzinsh's idea. Our adapted scheme enables the single-probe IFM of multiple objects with a linear scaling in resources using only the spatial degree of freedom of photons, at the cost of providing less complete information on the presence (or absence) of all interrogated objects.\par
Figure \ref{fig:two_objects} illustrates the approach for the case of two objects. Two EV interferometer units - as in Fig.\ref{fig:EV} - are arranged in cascade. The single photon enters the first EV interferometer to interrogate object 1. Similarly to the scheme in Fig.\ref{fig:Filatov}, the output port of the first interferometer that corresponds to an IFM success (the dark port $D_1$ in the standard EV setup) is fed into the input of a second interferometer that tests object 2. In other words, only if the photon emerges from the first stage having potentially “found” object 1 interaction-free, will it proceed to look for object 2. Meanwhile, unlike the Filatov-Auzinsh approach, here the light port is still connected to a photodetector, and not fed into a third EV unit overlapping with the second. At the output, the dark-port $D_2$ is monitored as the signature of an IFM of both objects. In this two-object chain, there are three main photon detection outcomes to consider (aside from absorption events):

\begin{enumerate}
    \item Top detector $D_0$ – the photon exited the first interferometer’s bright port without encountering object 1. In this case, it does not enter the second stage at all, and is simply detected at $D_0$ in our implementation. A click at $D_0$ thus provides no information about either object.
    \item Middle detector $D_1$ – the photon exited at the dark port of the first interferometer, but then went to the bright port of the second interferometer. In this scenario, we have evidence that object 1 is present but no information on object 2. A $D_1$ click thus indicates “object 1 is present, no information about object 2.”
    \item Bottom detector $D_{2}$ – the photon passed through the first interferometer’s dark port, meaning that object 1 has been detected, and then through the second interferometer’s dark port, detecting object 2 as well. A click at $D_2$ constitutes an interaction-free measurement of both objects simultaneously: it tells us that object 1 was in the first interferometer and object 2 was in the second, with the single photon surviving both encounters. This is the successful multi-object IFM outcome we are most interested in. In the ideal operation of this two-stage device, $D_2$ can never click unless both objects are actually present, since each stage’s dark port only yields a photon if its respective object is there to disrupt interference.
\end{enumerate}

The two-object scenario can be straightforwardly generalized to the $n$ object case. As represented in Fig.\ref{fig:n_objects}, for each additional object we add one more interferometer stage and one more output mode, with the IFM outcome being signalled by the lowermost detector. The generalized efficiency $\eta(n)$ is defined as the probability of the “all objects detected” outcome divided by the probability of either an IFM or an absorption occurring in the run:
\begin{equation}\label{eq:eta_multi_object}
\begin{aligned}
      \eta(n) & = \frac{P_{\rm IFM}(n)}{P_{\rm IFM}(n) + P_{\rm abs}^1 + ... + P_{\rm abs}^n} \\
      &
\end{aligned}
\end{equation}
where $P_{\rm IFM}(n)$ is the probability that the photon is detected in the final output $D_{n}$ - an interaction-free success for all $n$ objects - and $P_{\rm abs}^i$ the probability that the photon was absorbed by the $i^{th}$ object along the way. As we add more objects, $P_{\rm IFM}(n)$ tends to diminish and there will be a higher chance of an absorption by any of the objects occurring, so that $\eta(n)$ is in general a quickly decaying function of $n$. The probability of each outcome will in general depend on the reflectivities $R_i$ chosen for each pair of BSs. In App.\ref{appendix:simulations}, a detailed analysis of the choice of $R_i^{\text{opt}}$ which maximizes $\eta$ is provided. In summary, $R_i^{\text{opt}}$ is independent of $n$ and is as follows: $R_1^{\text{opt}}$ should be chosen as close to 1 as possible (just as in the standard EV protocol), while $R_i^{\text{opt}}, i=2,...,n$ asymptotically decrease towards $0.5$ for larger indices $i$. However, as pointed out in Sec.\ref{section:EV}, depending on practical considerations, other figures of merit might be more relevant for the multi-object protocol as for the standard EV one. For instance, if we adopt the choice of $R_i$ which maximizes the efficiency, $P_{\rm IFM}(n)$ becomes vanishingly small, especially with larger $n$, and the amount of time and resources required in order to complete an IFM of all objects becomes unfeasible. Therefore, for this type of protocol, a more realistic choice might be to fix all $R_i^{\text{opt}}$ closer to 0.5, the configuration that optimizes $P_{\rm IFM}(n)$, at the cost of having more absorption events.\par
We emphasize that our sequential approach does not reproduce all outcomes made possible by the fully symmetric scheme of \cite{filatovSetupInteractionfreeMeasurement2024}. In the two object scenario, we cannot extract the outcome corresponding to “object 2 present but not object 1”. If object 1 were absent but object 2 present, the photon would always exit at $D_0$ after stage 1, thus giving no info, and never even probe object 2 in stage 2. Thus, our scheme gives only partial information in cases where not all objects are present – essentially, it is biased to detect the earliest object in the sequence that is present, and it cannot detect later objects if an earlier one is missing. Despite this limitation, the sequential scheme still provides a valid demonstration of multi-object interaction-free detection: if all $n$ objects are present, there is a non-zero probability that the single photon will reveal this by propagating through all $n$ stages and exiting at the final dark port without being absorbed. The benefit of the sequential design is that it only requires $n+1$ modes to test $n$ objects, as opposed to $2^n$ distinct output paths in a fully overlapping design.\par
\begin{figure}
    \centering
    \includegraphics[width=.8\linewidth]{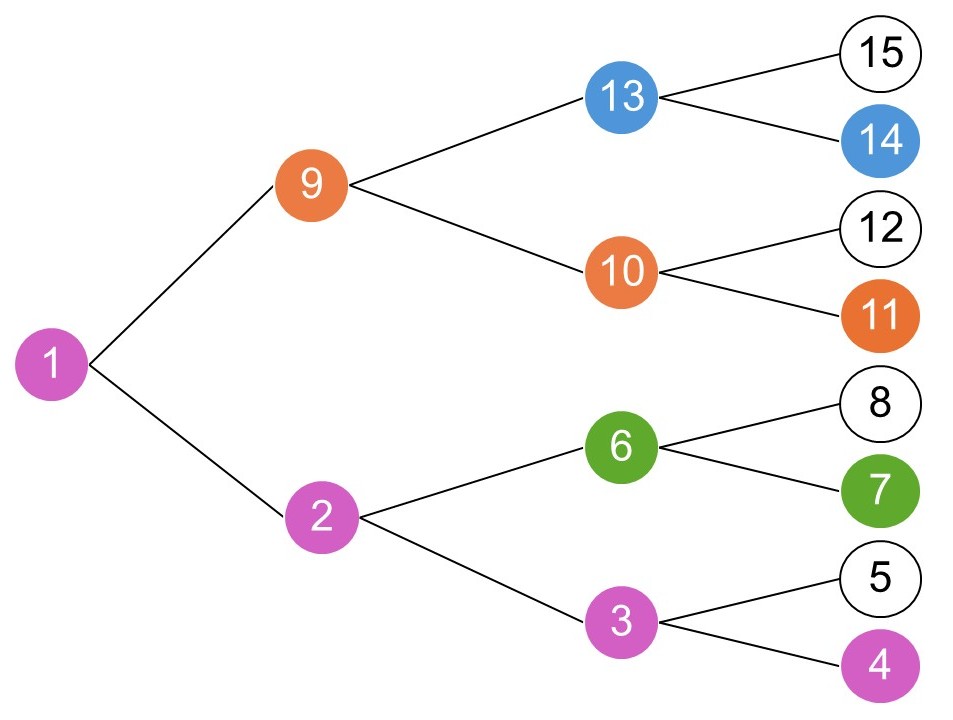}
    \caption{Extension of the scheme illustrated in Fig.\ref{fig:n_objects} for 15 objects, in a graph representation. Each node represents an EV unit, with edges representing input-output interferometer connections between them. A 16-mode interferometer can fit the quantum interrogation modules of 15 objects, in 4 layers of pairs of beamsplitters. Coloured sequences of nodes represent sets of objects interrogated by each unit which can be detected counterfactually with a single probe.}
    \label{fig:binary_tree}
\end{figure}

One extension of our proposal allows us to recycle a probe exiting towards the no-information outcome of a given EV unit for further interrogations, even without overlapping interferometers. Returning to the two-object case of Fig \ref{fig:two_objects}, one could introduce an additional optical mode above mode 0 and a third EV unit, possibly containing a third object, between these two modes, after the first EV unit. After interrogating the first object, the probe can exit towards the no-information outcome and interrogate the third object, or towards the IFM outcome, to interrogate the second object. A photon exiting the interferometer at the additional top mode would now indicate the presence of this third object, yet without providing any information on the other two. Output mode $0$ would correspond to the null outcome, with no information on the three objects. This scheme amounts to the one in Fig.\ref{fig:Filatov} but where instead of two overlapping interferometers interrogating the same object, we have a linear structure of one EV unit followed by two non-overlapping devices interrogating two different objects. Representing each EV unit, possibly containing object $i$, as a node in a graph, we can represent this scheme as a binary tree, where the edges represent output-to-input connection between the interferometers. This is shown in Fig.\ref{fig:binary_tree}, for a setup with 15 objects arranged in 4 layers. Each coloured, descending linear chain represents a sequence of objects that can be detected with no interaction with a single probe. Uncoloured nodes represent objects that can still be detected using recycled probes that exited the light port in all previous EV units. In general, $\sum_{i=0}^{k-1}2^i=2^k -1$ objects arranged in $k$ vertical layers can be arranged in a $2^k$-mode linear optical circuit with depth $2k$. There will be a single chain of $k$ objects that can be interrogated with the same probe, as well as a chain of $k-1$ objects, 2 chains of $k-2$ objects, 3 of $k-3$ objects, and so on. This extension showcases the flexibility of our proposal, which enables different configurations of multiple objects resulting in different sets of partial information that can be extracted.\par
Finally, as is true for the Filatov-Auzinsh scheme \cite{filatovSetupInteractionfreeMeasurement2024}, each EV box in Fig.\ref{fig:n_objects} could in principle be replaced with the high efficiency Kwiat setup \cite{kwiatInteractionFreeMeasurement1995} for definite information even in the case of a light port outcome and for improved efficiency. One could also envision using, in place of the EV units, the scheme represented in Fig.\ref{fig:UUdagger} for possibly reduced noise in photon detection events in a setting with independent errors in optical components, as discussed in App.\ref{appendix:robustness}, or even the one proposed in \cite{paliciInteractionfreeImagingMultipixel2022} in order to interrogate multi-pixel objects. 

\section{Experimental implementation and results}\label{section:exp}

In this work, we experimentally implemented two IFM tasks using the cloud-accessible Ascella Quantum Processing Unit (QPU), provided by the Quandela Cloud service \cite{maringVersatileSinglephotonbasedQuantum2024}. A universal linear-optical transformation on $m$ modes requires $m(m-1)/2$ variable beam-splitters and phase shifters. The Ascella QPU consists of a 12-mode programmable universal interferometer \cite{reckExperimentalRealizationAny1994,clementsOptimalDesignUniversal2016}, a rectangular mesh of balanced beamsplitters and tunable phase shifters arranged in a grid of MZI units, which effectively act as beamsplitters of variable reflectivity, interspersed with phase shifters.  By tuning the phase shifters, the interferometer can be reconfigured to implement a generic 12-mode linear-optical mapping of input to output creation operators. It is equipped with an on-demand quantum dot single-photon source and superconducting nanowire single-photon detectors at each output (see App.\ref{sec:methods} for more details). In Sec.\ref{section:EV_UPP} we benchmark the performance of the Ascella QPU by implementing the well-established EV quantum interrogation task. This benchmarking validates the results presented in Sec.\ref{section:multiple_object_UPP}, where we report the implementation of our proposal for a multiple-object IFM scheme, detailed in Sec.\ref{section:our_multiple_object}. \par
The architecture of an integrated photonic processor does not allow the introduction of absorbers inside the spatial modes of the device. To emulate the presence of absorptive objects within the interferometer, we adopted the approach used in Ref.\cite{giordaniExperimentalCertificationContextuality2023}. Namely, an “object” blocking a given optical path is modelled by diverting that path to an output port connected to a dedicated single-photon detector, not allowing it to interfere with other paths. A click in that detector then signifies that the photon would have been absorbed by the object. By contrast, if the photon is not routed to this absorbing detector, it continues through the interferometer and can contribute to an interaction-free detection event. Using this technique, we can effectively insert or remove absorptive objects in different arms of the interferometer by reprogramming the linear optical circuit.\par 
In order to verify that observed IFM events, as well as absorption events in the object detectors, were in fact due to the presence of obstruction, and not simply due to noise in the implemented setups, we conducted, for each reported setup, an equivalent experiment where the mode permutations redirecting the photon to object detectors were not introduced. App.\ref{appendix:data_without_object} discusses a comparison between the data obtained with and without object obstruction, which support the validity of our experimental results.

\subsection{The standard quantum interrogation task}\label{section:EV_UPP}

\begin{figure*}
    \centering
    \subfigure[]{
       \includegraphics[width=0.475\textwidth,valign = m]{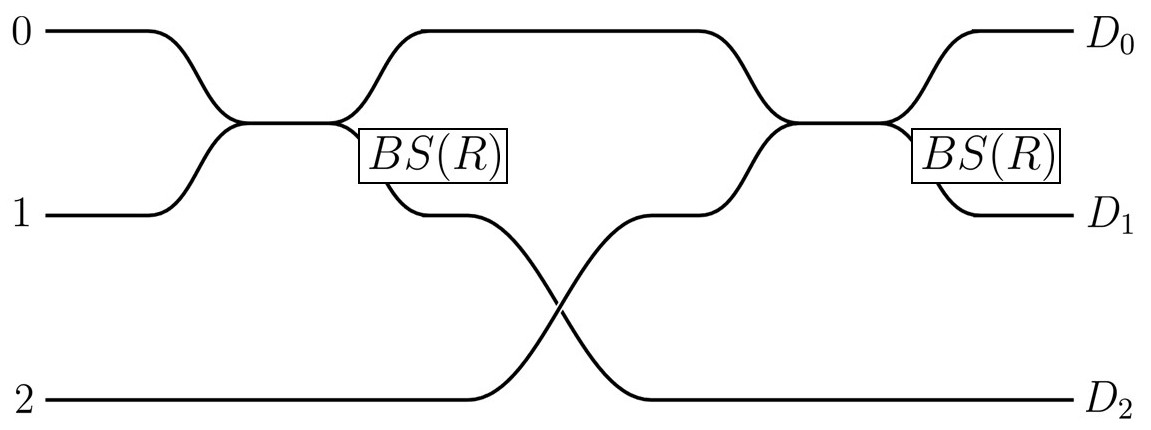}\label{fig:EV_scheme}}
    \subfigure[]{
        \includegraphics[width=0.5\textwidth,valign = m]{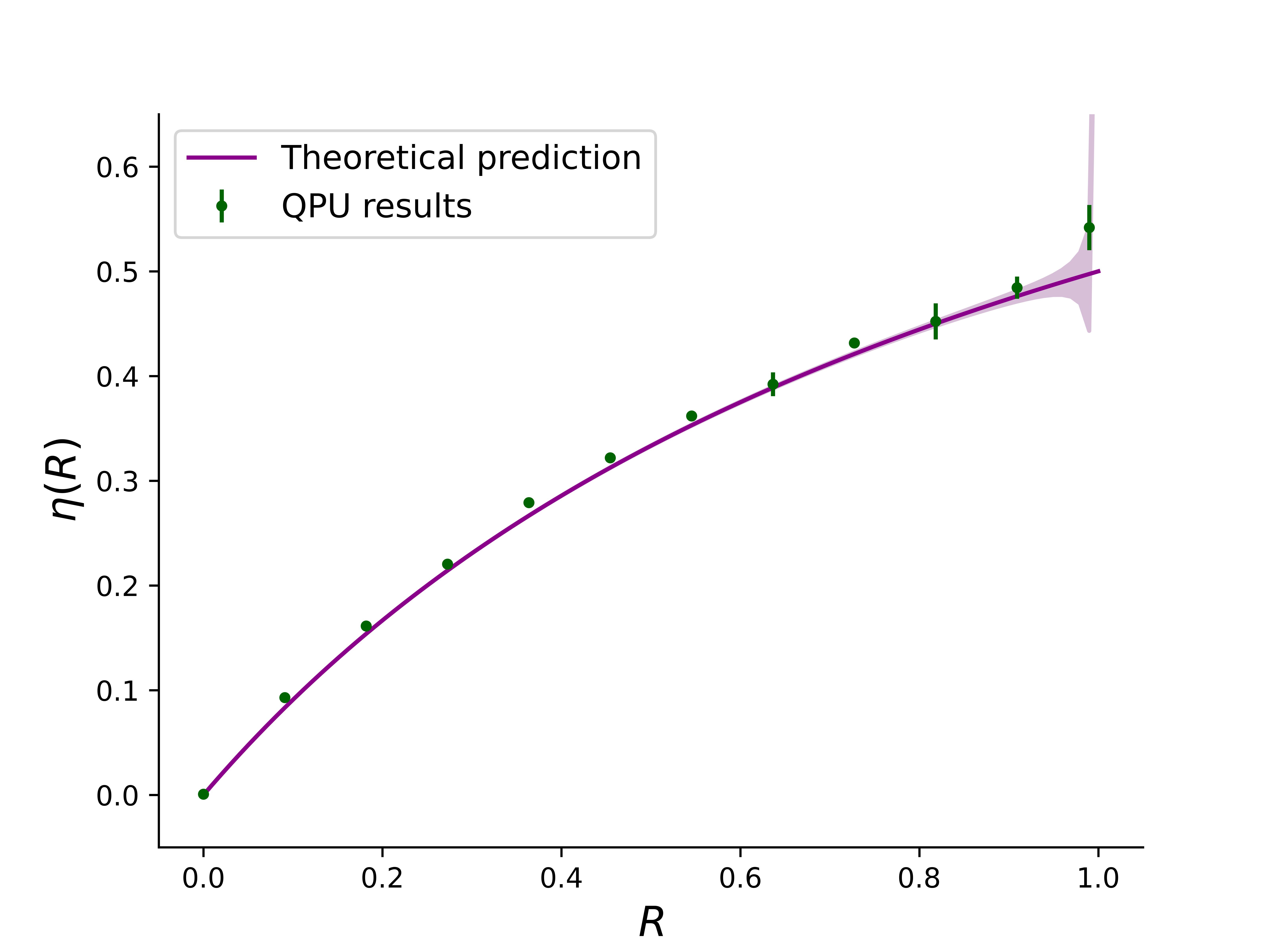}  \label{fig:eta(R)_plot}}
    \caption{(a) Circuit implemented in Perceval for the standard Elitzur-Vaidman protocol. The opaque object is modelled by photodetector $D_2$, which heralds the absorption of the photon. (b) Efficiency $\eta$ of the standard Elitzur-Vaidman protocol for an interaction-free measurement as a function of beam splitter reflectivity $R$, obtained from sampling in Quandela's quantum processing unit Ascella. The data closely follows the analytical curve in magenta, given by eq.\eqref{eq:eta_EV}. The violet shaded area captures deviations due to imperfect calibration of the device. Error bars are obtained from an error mitigation technique (see details in App.\ref{appendix:errorbars}).}
\end{figure*}

We implement the scheme of Fig.\ref{fig:EV} with variable $R$ using the photonic processor. The presence of the object is emulated by diverting the obstructed optical path to a third photodetector $D_2$, as can be seen in Fig.\ref{fig:EV_scheme}. If the photon takes path 1 after the BS, it is redirected towards $D_2$, so that detection in this output mode corresponds to absorption of the photon by the object. The detector itself can be thought of as the absorbing object that heralds the absorption. The results are reported in Fig.\ref{fig:eta(R)_plot}. In the experiment, we vary the reflectivity $R$ of the BSs in the range $R\in\, ]0,1[$ and measure the resulting probabilities of each output detector clicking. Of particular interest is the dark-port detector $D_1$, which in the ideal case clicks only if the object is present (signalling a successful interaction-free detection), and detector $D_2$ modelling an absorption event by the opaque object. From these, we calculate the IFM efficiency $\eta(R)$ for each setting, as given by eq. \eqref{eq:eta_EV}. Figure \ref{fig:eta(R)_plot} shows the measured efficiency - green data points - as a function of $R$, together with the theoretical curve in magenta, expected from Eq.\eqref{eq:eta_EV_R}. A main source of noise affecting our data is imperfections in optical components of the chip, which are compensated for by a global, machine-learning-based calibration technique tailored to the QPU \cite{fyrillasScalableMachineLearningassisted2024}. Furthermore, we found that compilation/translation artefacts associated to this procedure introduced a systematic bias for some of the implemented unitaries. While for most values of $R$, the measured $\eta(R)$ was in good agreement to the theoretical curve, for certain values, estimates of the efficiency obtained in different experiments consistently exhibited a significant deviation from the theoretical value, as large as $40\%$ relative error in some cases. We attributed this systematic deviation to a poor convergence of the calibration routine, and adopted an error mitigation technique based on ensemble averaging to overcome this unwanted effect. Efficiency estimates were extracted and averaged over multiple equivalent circuits, differing only in physically irrelevant phases that corresponded to distinct initial configuration parameters fed to the compilation algorithm. This helped the iteration procedure to better converge, greatly enhancing the fidelity of implemented unitaries and leading to better agreement with the expected results. More details on these methods, as well as a comparison between the raw and error-mitigated data illustrating the sensitivity of $\eta(R)$ to the mitigation procedure, can be found in App.\ref{appendix:errorbars}.\par
As $R\to 1$, the efficiency is predicted to approach $50\%$ asymptotically. Our data reflect this rise in $\eta$ for large $R$, but with an increasing divergence from the ideal curve at the extreme. We attribute this discrepancy to an imperfect on-chip implementation of BSs that are nearly perfectly reflecting. In practice, setting a very large reflectivity in the compiled circuit can be sensitive to slight mismatches – for example, if the reflectivity of the first BS $R$ does not exactly match the second BS’s reflectivity $R'$, the two interferometer arms are not perfectly balanced, and the dark port is not truly dark even without an object. This leads to additional leakage of photons to $D_1$ - or equivalently, less-than-expected destructive interference - skewing the efficiency calculation. Indeed, as was pointed out in \cite{giordaniExperimentalCertificationContextuality2023}, this error becomes more significant as $R\to1$, since in this regime both $P_{\rm IFM}$ and $P_{\rm abs}$ are very small, so any imperfections or background counts can cause large relative fluctuations in $\eta$. Following \cite{giordaniExperimentalCertificationContextuality2023}, we can write a noise model for $\eta$ in terms of the relative mismatch 

\begin{equation}
    \varepsilon = \frac{|R-R'|}{R}
\end{equation}

\noindent between the reflectivities

\begin{equation}
    \eta(\varepsilon) = \frac{R[1-(1\pm\varepsilon)R]}{R[1-(1\pm\varepsilon)R] - R + 1}
\end{equation}
We represent this deviation from the ideal curve as a shaded area in Fig.\ref{fig:eta(R)_plot}(b), fixing $\varepsilon = 0.2\%$. We see that a reflectivity mismatch of this magnitude is enough to explain the deviations from the curve towards higher $R$ values. In App.\ref{appendix:robustness}, we present simulated data that corroborates this higher sensitivity as we approach the efficiency ceiling. We also discuss how setups with higher number of modes as described in Sec.\ref{section:multimode} could help mitigate this sensitivity in $R$ errors. However, this last observation only holds for physical systems where errors in optical components can be considered independent. This is not the case in the UPP exploited in this work, where all optical components are simultaneously configured in a global optimization procedure, leading to correlated errors (see App.\ref{appendix:robustness} for more details).\par 
Other sources of error, such as detector dark counts or errors in the chip configuration, are inaccessible due to the lack of direct control over the experimental apparatus. Nevertheless, the single-object IFM behaviour observed on the chip agrees well with theory, confirming that the photonic processor can faithfully reproduce the EV interaction-free measurement. 

\subsection{Interrogation of multiple objects}\label{section:multiple_object_UPP}

\begin{figure*}
    \centering
    \subfigure[]{
    \includegraphics[width=.6\textwidth]{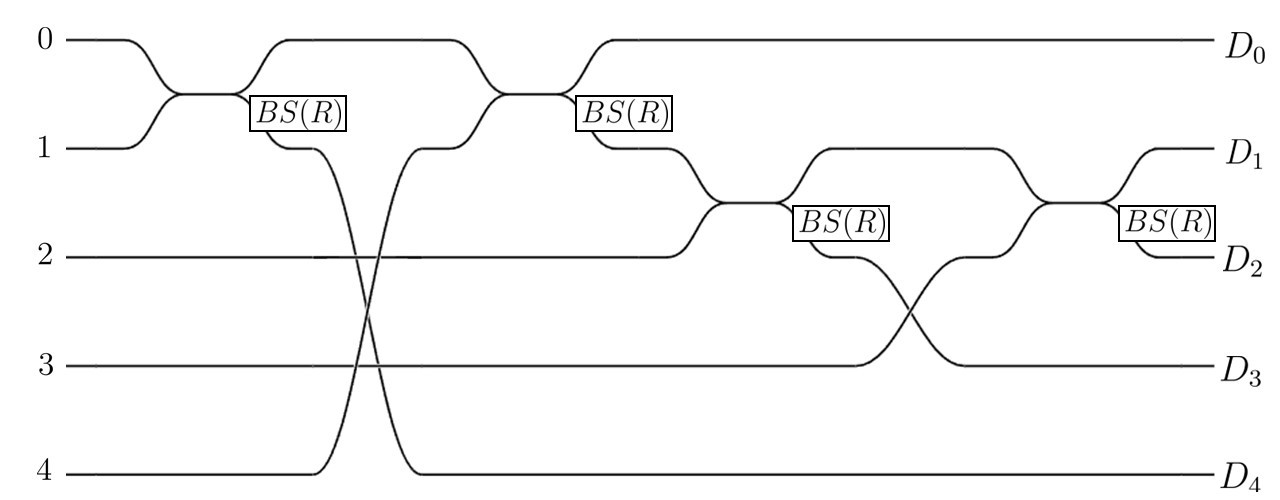}
    \label{fig:two_bomb_circuit}}
    \subfigure[]{
    \includegraphics[width=0.5\textwidth]{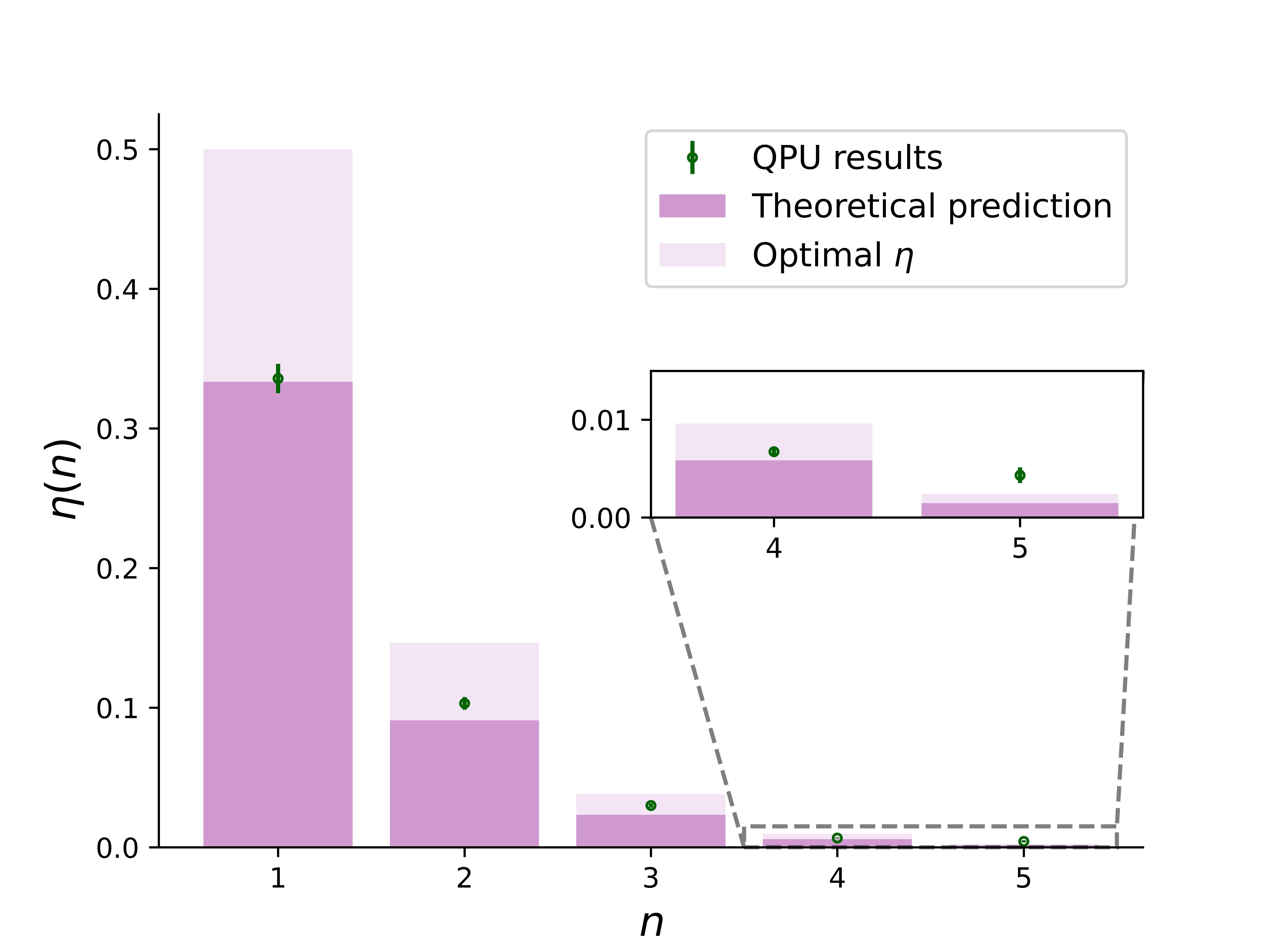}
    \label{fig:eta(n)_plot}}
    \caption{(a) Scheme of the circuit used to implement a  single-probe interaction-free measurement of two objects. (b) Efficiency $\eta(n)$ of the single-probe interaction-free measurement of $n$ objects. The experimental results from the QPU, in green, were obtained for a beamsplitter reflectivity of $R=0.5$ for all pairs of beam splitters in the generalized scheme, and follow the theoretical prediction in dark pink. We compared these to the numerically computed optimal results in light pink, obtained for a different choice of $R$ (see App.\ref{appendix:simulations}). Error bars are obtained from an error mitigation technique (see details in App.\ref{appendix:errorbars}).}
\end{figure*}

The UPP's architecture cannot directly realize the fully overlapping interferometer configuration of Ref \cite{filatovSetupInteractionfreeMeasurement2024}, nor the manipulation of the time degree of freedom of the photons, making it an unsuitable platform for experimental implementation of the Filatov-Auzinsh proposal. The Ascella device is essentially a fixed mesh of beam splitters – we can route modes in various ways, but we cannot easily overlap two distinct interferometers in the same set of modes without crosstalk, nor can we dynamically reuse the same photon in multiple passes through the chip, as we do not have optical storage or time-delay loops integrated. In contrast, our simplified protocol introduced in Sec.\ref{section:our_multiple_object} enables the realization of the generalized IFM of multiple-objects, within the limitations of the available hardware. The linear scaling in resource usage made it feasible to implement up to $n=5$ objects on the 12-mode chip we exploited.\par
Fig.\ref{fig:eta(n)_plot} plots the resulting efficiency for each $n$, given by eq. \eqref{eq:eta_multi_object}, from 1 up to 5 - green circles - alongside the bars of expected theoretical values. For each $n$, we implemented a circuit such as the one exemplified in Fig.\ref{fig:two_bomb_circuit} for the case of two objects, where absorption by each object is modelled by redirection to a photodetector. The reflectivities of each pair of BSs in the EV units were all set to $R_i=0.5$. This simplification leads to the following dependence of the efficiency on the number of objects considered:
\begin{equation}
    \eta(n)_{R_i=0.5} = \frac{1}{1 + \frac{2}{3}(4^n-1)}
\end{equation}
This does not correspond to the optimal configuration described in Sec.\ref{section:our_multiple_object} that yields the maximum efficiency, represented in the graph in light pink bars for comparison. This choice stems primarily from experimental limitations, namely the difficulty of accumulating photon-counting statistics in the optimal configuration. As mentioned in Sec.~\ref{section:our_multiple_object}, the optimal configuration corresponds to the limit of $R_1\to 1$  for the first pair of BSs, while the remaining $R_i^{\text{opt}}$ are set to fixed values asymptotically approaching $0.5$ as $i$ increases. In this limit, $P_{\rm IFM}(n)$ becomes vanishingly small, particularly for schemes with many objects, so its estimation would require a large volume of statistics, hard to obtain experimentally on the cloud-accessible QPU. We therefore opted for the configuration that maximizes $P_{\rm IFM}(n)$, which still yields an efficiency close to the optimal value for larger $n$. Further motivation is provided in App.~\ref{appendix:simulations}.\par
We obtained $P_{\rm IFM}(n)$ and the $P^{i}_{\rm abs}$ directly from the measured frequencies of $D_{n}$ clicks and each object’s absorption detector clicks, respectively. Similarly to what was reported in Sec.\ref{section:EV_UPP} for the EV scheme, systematic errors due to compilation artefacts during the QPU's calibration procedure were found to significantly affect the efficiency estimates for the multi-object setups. Thus, the error mitigation technique validated for the standard EV experiment was also applied to the data from the experiments reported in this section, leading to a greater accuracy in obtained results. A comparison between the raw and treated data can be found in App. \ref{appendix:errorbars}. Despite the higher complexity of the sequential process, the experimental data align well with the predicted $\eta(n)_{R_i=0.5}$ for all tested values. Notably, for $n=5$ we clearly observe a non-zero efficiency - on the order of $\eta(5)_{R_i=0.5}\approx 0.1\%$ - meaning the chip succeeded in performing a counterfactual measurement of five separate objects in one go. To our knowledge this is the first time an interaction-free measurement involving more than one object has been realized experimentally. The efficiency does diminish rapidly as $n$ increases, which is expected because the probability of the photon surviving all encounters falls off (roughly exponentially) with $n$, even in the optimal efficiency scenario. This behaviour underscores the need for more sophisticated schemes if one hopes to scale up to many objects. For example, as already mentioned in Sec.\ref{section:our_multiple_object}, one could incorporate the quantum Zeno-based high-efficiency modules at each stage as suggested in \cite{filatovSetupInteractionfreeMeasurement2024}: if each stage had an efficiency near $100\%$, then even a large cascade could maintain a high overall success probability. The hardware constraints of the UPP do not enable the implementation of this more complex scheme, but the framework we established could be extended by replacing each EV interferometer with a multi-pass equivalent.\par

\section{Discussion}\label{section:discussion}

Integrated photonics platforms have brought significant developments to the implementation of photonic quantum information tasks, offering miniaturization, stability and versatility. Cloud-based services such as Quandela Cloud are emerging as a new, practical tool for theoreticians to design and test proof-of-principle demonstrations and reproduce experimental results without needing to have direct access to an experimental setup. In this work, we have used this tool for an implementation of the landmark Elitzur-Vaidman quantum interrogation task on Quandela's Ascella universal photonic processor. Taking advantage of the reconfigurable, multimode architecture of the chip, which seamlessly enables the introduction of multiple absorbers emulated as redirections to photodetectors, we have demonstrated what is, to the best of our knowledge, the first experimental realization of a generalized interaction-free measurement protocol for multiple objects using a single probe. Our proof-of-principle experiments confirm that a single photon can be used to sequentially probe multiple spatially separated objects and detect their presence in a single click without interacting with them. This work goes beyond earlier proposals such as the one by Filatov and Auzinsh for the interrogation of multiple objects, by implementing such a scheme on real hardware and verifying its operation with up to five emulated objects. The efficiency of our proposed protocol is modest, owing to the quickly decaying probability that the probe avoids all objects and reaches the IFM detector as the number of objects increases, but these and other relevant figures of merit could be improved by leveraging known high-efficiency IFM techniques \cite{kwiatExperimentalTheoreticalProgress1998}. Advancements in integrated photonic hardware could enable implementation of the larger, low-loss circuits required for these more intricate schemes. In future work, potential applications of our scheme to interaction-free quantum imaging schemes could be explored. By counterfactually imaging multiple objects, or multiple pixels in a sample, in a single shot, this scheme could simultaneously reduce exposure of the samples to light compared to conventional imaging methods while increasing the speed of the imaging process compared to single-object IFM. Overall, our results highlight the power of integrated photonics as a versatile test-bed for quantum measurement concepts and mark an important step in generalizing interaction-free measurements to more complex scenarios.

\section*{Code availability}\label{sec:code}

Code used for reproducing the experimental error-mitigated results in this work using the Perceval framework, as well as files containing the data obtained from the QPU, can be found at \href{https://github.com/sara-rdf/Bomb-testing-on-a-photonic-processor}{https://github.com/sara-rdf/Bomb-testing-on-a-photonic-processor}.

\section*{Acknowledgements}

The authors thank Pierre-Emmanuel Emeriau, Emilio Annoni and Rawad Mezher for advice on error characterization and mitigation, and the team at Quandela for fruitful discussions and for making the Perceval framework and the Quandela Cloud service available to the scientific community. SF acknowledges support from ERC Advanced grant QU-BOSS (GA No. 884676) and from FCT–Fundação para a Ciência e a Tecnologia (Portugal) through PhD grant 2025.07305.BDANA. A.C. acknowledges financial support from FCT - Fundação para a Ciência e a Tecnologia (Portugal) via PhD Grant SFRH/BD/151190/2021. EFG acknowledges support from FCT–Fundação para a Ciência e a Tecnologia (Portugal) via project CEECINST/00062/2018, and from the National Council for
Scientific and Technological Development – CNPq (Brazil) under grant 308292/2025-1.

\bibliographystyle{unsrt}

\bibliography{references}

@article{kentQuantumInterrogationSafer2001,
	title = {Quantum {Interrogation} and the {Safer} {X}-ray},
	journal = {arXiv:quant-ph/0102118},
	author = {Kent, Adrian and Wallace, David},
	year = {2001},
}

@article{hanceCounterfactualGhostImaging2021,
	title = {Counterfactual ghost imaging},
	volume = {7},
	issn = {20566387},
	doi = {10.1038/s41534-021-00411-4},
	number = {1},
	journal = {npj Quantum Inf.},
	publisher = {Nature Research},
	author = {Hance, Jonte R. and Rarity, John},
	year = {2021},
	pages = {88},
}

@article{renningerMessungenOhneStorung1960,
	title = {Messungen ohne {Storung} des {Messobjekts} ({Measurement} without disturbance of the measured objects)},
	volume = {158},
	doi = {10.1007/BF01327019},
	journal = {Z. Physik},
	author = {Renninger, Mauritius},
	year = {1960},
	pages = {417--421},
}

@article{clementsOptimalDesignUniversal2016,
	title = {Optimal {Design} for {Universal} {Multiport} {Interferometers}},
	volume = {3},
	doi = {10.1364/OPTICA.3.001460},
	language = {en},
	journal = {Optica},
	publisher = {arXiv},
	author = {Clements, William R. and Humphreys, Peter C. and Metcalf, Benjamin J. and Kolthammer, W. Steven and Walmsley, Ian A.},
	year = {2016},
	pages = {1460--1465},
}

@article{paliciInteractionfreeImagingMultipixel2022,
	title = {Interaction-free imaging of multipixel objects},
	volume = {105},
	issn = {2469-9926, 2469-9934},
	url = {https://link.aps.org/doi/10.1103/PhysRevA.105.013529},
	doi = {10.1103/PhysRevA.105.013529},
	language = {en},
	number = {1},
	journal = {Phys. Rev. A},
	author = {Pălici, Alexandra Maria and Isdrailă, Tudor-Alexandru and Ataman, Stefan and Ionicioiu, Radu},
	year = {2022},
	pages = {013529},
}

@article{hafnerExperimentInteractionfreeMeasurement1997,
	title = {Experiment on interaction-free measurement in neutron interferometry},
	volume = {235},
	copyright = {https://www.elsevier.com/tdm/userlicense/1.0/},
	issn = {03759601},
	url = {https://linkinghub.elsevier.com/retrieve/pii/S0375960197006968},
	doi = {10.1016/S0375-9601(97)00696-8},
	language = {en},
	number = {6},
	journal = {Phys. Rev. A},
	author = {Hafner, Meinrad and Summhammer, Johann},
	year = {1997},
	pages = {563--568},
}

@article{dumarchievanvoorthuysenRealizationInteractionfreeMeasurement1996,
	title = {Realization of an interaction-free measurement of the presence of an object in a light beam},
	volume = {64},
	issn = {0002-9505, 1943-2909},
	url = {https://pubs.aip.org/ajp/article/64/12/1504/1054750/Realization-of-an-interaction-free-measurement-of},
	doi = {10.1119/1.18413},
	language = {en},
	number = {12},
	journal = {Am. J. Phys.},
	author = {Du Marchie Van Voorthuysen, E. H.},
	year = {1996},
	pages = {1504--1507},
}

@article{filatovSetupInteractionfreeMeasurement2024,
	title = {Setup for interaction-free measurement of multiple objects using single quantum probe},
	volume = {130},
	issn = {0946-2171, 1432-0649},
	url = {https://link.springer.com/10.1007/s00340-024-08257-2},
	doi = {10.1007/s00340-024-08257-2},
	language = {en},
	number = {7},
	journal = {Appl. Phys. B},
	author = {Filatov, Stanislav and Auzinsh, Marcis},
	year = {2024},
	pages = {121},
}

@article{caoCounterfactualUniversalQuantum2020,
	title = {Counterfactual universal quantum computation},
	volume = {102},
	issn = {2469-9926, 2469-9934},
	url = {http://arxiv.org/abs/2011.07195},
	doi = {10.1103/PhysRevA.102.052413},
	language = {en},
	number = {5},
	journal = {Phys. Rev. A},
	author = {Cao, Zhu},
	year = {2020},
	pages = {052413},
}

@article{salihDeterministicTeleportationUniversal2025,
	title = {Deterministic {Teleportation} and {Universal} {Computation} {Without} {Particle} {Exchange}},
	volume = {9},
	url = {http://arxiv.org/abs/2009.05564},
	doi = {10.1088/2399-6528/adce54},
	journal = {J. Phys. Commun.},
	author = {Salih, Hatim and Hance, Jonte R. and McCutcheon, Will and Rudolph, Terry and Rarity, John},
	year = {2025},
	pages = {045004},
}

@article{salihExchangefreeComputationUnknown2021,
	title = {Exchange-free computation on an unknown qubit at a distance},
	volume = {23},
	issn = {13672630},
	doi = {10.1088/1367-2630/abd3c4},
	number = {1},
	journal = {New Journal of Physics},
	publisher = {IOP Publishing Ltd},
	author = {Salih, Hatim and Hance, Jonte R. and McCutcheon, Will and Rudolph, Terry and Rarity, John},
	year = {2021},
	pages = {013004},
}

@article{yangInteractionfreeSinglepixelQuantum2023,
	title = {Interaction-free, single-pixel quantum imaging with undetected photons},
	volume = {9},
	issn = {20566387},
	doi = {10.1038/s41534-022-00673-6},
	number = {1},
	journal = {npj Quantum Inf},
	publisher = {Nature Research},
	author = {Yang, Yiquan and Liang, Hong and Xu, Xiaze and Zhang, Lijian and Zhu, Shining and Ma, Xiao song},
	year = {2023},
	pages = {2},
}

@article{liuExperimentalDemonstrationCounterfactual2012,
	title = {Experimental demonstration of counterfactual quantum communication},
	volume = {109},
	issn = {0031-9007, 1079-7114},
	url = {http://arxiv.org/abs/1107.5754},
	doi = {10.1103/PhysRevLett.109.030501},
	language = {en},
	number = {3},
	journal = {Phys. Rev. Lett.},
	author = {Liu, Yang and Ju, Lei and Liang, Xiao-Lei and Tang, Shi-Biao and Tu, Guo-Liang Shen and Zhou, Lei and Peng, Cheng-Zhi and Chen, Kai and Chen, Teng-Yun and Chen, Zeng-Bing and Pan, Jian-Wei},
	year = {2012},
	pages = {030501},
}

@article{fyrillasScalableMachineLearningassisted2024,
	title = {Scalable machine learning-assisted clear-box characterization for optimally controlled photonic circuits},
	volume = {11},
	issn = {2334-2536},
	url = {https://opg.optica.org/abstract.cfm?URI=optica-11-3-427},
	doi = {10.1364/OPTICA.512148},
	language = {en},
	number = {3},
	journal = {Optica},
	author = {Fyrillas, Andreas and Faure, Olivier and Maring, Nicolas and Senellart, Jean and Belabas, Nadia},
	year = {2024},
	pages = {427},
}

@article{guoQuantumCryptographyBased1999,
	title = {Quantum cryptography based on interaction-free measurement},
	volume = {256},
	copyright = {https://www.elsevier.com/tdm/userlicense/1.0/},
	issn = {03759601},
	url = {https://linkinghub.elsevier.com/retrieve/pii/S0375960199002352},
	doi = {10.1016/S0375-9601(99)00235-2},
	language = {en},
	number = {2-3},
	journal = {Physics Letters A},
	author = {Guo, Guang-Can and Shi, Bao-Sen},
	year = {1999},
	pages = {109--112},
}

@article{kwiatInteractionFreeMeasurement1995,
	title = {Interaction-{Free} {Measurement}},
	volume = {74},
	doi = {10.1103/PhysRevLett.74.4763},
	number = {24},
	journal = {Phys. Rev. Lett.},
	author = {Kwiat, Paul and Weinfurter, Harald and Herzog, Thomas and Zeilinger, Anton},
	year = {1995},
	pages = {4763--4766},
}

@article{maOnchipInteractionfreeMeasurements2014,
	title = {On-chip interaction-free measurements via the quantum {Zeno} effect},
	volume = {90},
	url = {http://arxiv.org/abs/1405.2068},
	doi = {10.1103/PhysRevA.90.042109},
	number = {4},
	journal = {Phys. Rev. A},
	author = {Ma, Xiao-Song and Guo, Xiang and Schuck, Carsten and Fong, King Y. and Jiang, Liang and Tang, Hong X.},
	year = {2014},
	pages = {042109},
}

@article{heurtelPercevalSoftwarePlatform2023,
	title = {Perceval: {A} {Software} {Platform} for {Discrete} {Variable} {Photonic} {Quantum} {Computing}},
	volume = {7},
	issn = {2521-327X},
	shorttitle = {Perceval},
	doi = {10.22331/q-2023-02-21-931},
	language = {en},
	journal = {Quantum},
	author = {Heurtel, Nicolas and Fyrillas, Andreas and Gliniasty, Grégoire de and Bihan, Raphaël Le and Malherbe, Sébastien and Pailhas, Marceau and Bertasi, Eric and Bourdoncle, Boris and Emeriau, Pierre-Emmanuel and Mezher, Rawad and Music, Luka and Belabas, Nadia and Valiron, Benoît and Senellart, Pascale and Mansfield, Shane and Senellart, Jean},
	year = {2023},
	pages = {931},
}

@article{maringVersatileSinglephotonbasedQuantum2024,
	title = {A versatile single-photon-based quantum computing platform},
	volume = {18},
	url = {https://www.nature.com/articles/s41566-024-01403-4},
	doi = {10.1038/s41566-024-01403-4},
	language = {en},
	number = {6},
	journal = {Nature Photonics},
	author = {Maring, Nicolas and Fyrillas, Andreas and Pont, Mathias and Ivanov, Edouard and Stepanov, Petr and Margaria, Nico and Hease, William and Pishchagin, Anton and Lemaître, Aristide and Sagnes, Isabelle and Au, Thi Huong and Boissier, Sébastien and Bertasi, Eric and Baert, Aurélien and Valdivia, Mario and Billard, Marie and Acar, Ozan and Brieussel, Alexandre and Mezher, Rawad and Wein, Stephen C. and Salavrakos, Alexia and Sinnott, Patrick and Fioretto, Dario A. and Emeriau, Pierre-Emmanuel and Belabas, Nadia and Mansfield, Shane and Senellart, Pascale and Senellart, Jean and Somaschi, Niccolo},
	year = {2024},
	pages = {603--609},
}

@article{giordaniExperimentalCertificationContextuality2023,
	title = {Experimental certification of contextuality, coherence, and dimension in a programmable universal photonic processor},
	volume = {9},
	url = {https://www.science.org},
	doi = {10.1126/sciadv.adj4249},
	journal = {Sci. Adv.},
	author = {Giordani, Taira and Wagner, Rafael and Esposito, Chiara and Camillini, Anita and Hoch, Francesco and Carvacho, Gonzalo and Pentangelo, Ciro and Ceccarelli, Francesco and Piacentini, Simone and Crespi, Andrea and Spagnolo, Nicolò and Osellame, Roberto and Galvão, Ernesto F and Sciarrino, Fabio},
	year = {2023},
	pages = {eadj4249},
}

@article{kwiatHighefficiencyQuantumInterrogation1999,
	title = {High-efficiency quantum interrogation measurements via the quantum {Zeno} effect},
	volume = {833},
	doi = {10.1103/PhysRevLett.83.4725},
	number = {23},
	journal = {Phys. Rev. Lett.},
	author = {Kwiat, Paul and White, A. G. and Mitchell, J. R. and Nairz, O. and Weihs, G. and Weinfurter, H. and Zeilinger, Anton},
	year = {1999},
	pages = {4725--4728},
}

@article{kwiatExperimentalTheoreticalProgress1998,
	title = {Experimental and {Theoretical} {Progress} in {Interaction}-free {Measurements}},
	volume = {76},
	doi = {10.1238/Physica.Topical.076a00115},
	journal = {Physica Scripta},
	author = {Kwiat, Paul},
	year = {1998},
	pages = {115--121},
}

@article{wangIntegratedPhotonicQuantum2020,
	title = {Integrated photonic quantum technologies},
	volume = {14},
	doi = {https://doi.org/10.1038/s41566-019-0532-1},
	journal = {Nat. Photonics},
	author = {Wang, J. and Sciarrino, Fabio and Laing, A. and Thompson, Mark G.},
	year = {2020},
	pages = {273--284},
}

@article{metcalfMultiphotonQuantumInterference2013,
	title = {Multiphoton quantum interference in a multiport integrated photonic device},
	volume = {4},
	issn = {2041-1723},
	url = {https://doi.org/10.1038/ncomms2349},
	doi = {10.1038/ncomms2349},
	number = {1},
	journal = {Nature Communications},
	author = {Metcalf, Benjamin J. and Thomas-Peter, Nicholas and Spring, Justin B. and Kundys, Dmytro and Broome, Matthew A. and Humphreys, Peter C. and Jin, Xian-Min and Barbieri, Marco and Steven Kolthammer, W. and Gates, James C. and Smith, Brian J. and Langford, Nathan K. and Smith, Peter G.R. and Walmsley, Ian A.},
	year = {2013},
	pages = {1356},
}

@article{itanoQuantumZenoEffect1990,
	title = {Quantum {Zeno} effect},
	volume = {41},
	doi = {10.1103/PhysRevA.41.2295},
	number = {5},
	journal = {Phys. Rev. A},
	author = {Itano, Wayne M. and Heinzen, D. J. and Bollinger, J. J. and Wineland, D. J.},
	year = {1990},
	pages = {2295--2300},
}

@article{misraZenosParadoxQuantum1977,
	title = {The {Zeno}'s paradox in quantum theory},
	volume = {18},
	doi = {https://doi.org/10.1063/1.523304},
	journal = {J. Math. Phys.},
	author = {Misra, B. and Sudarshan, E. C. G.},
	year = {1977},
	pages = {756--763},
}

@article{reckExperimentalRealizationAny1994,
	title = {Experimental realization of any discrete unitary operator},
	volume = {73},
	doi = {10.1103/PhysRevLett.73.58},
	number = {1},
	journal = {Phys. Rev. Lett.},
	author = {Reck, Michael and Zeilinger, Anton and Bernstein, Herbert J. and Bertani, Philip},
	year = {1994},
	pages = {58--61},
}

@article{nohCounterfactualQuantumCryptography2009,
	title = {Counterfactual {Quantum} {Cryptography}},
	volume = {103},
	doi = {10.1103/PhysRevLett.103.230501},
	language = {en},
	number = {23},
	journal = {Phys. Rev. Lett.},
	author = {Noh, Tae-Gon},
	year = {2009},
	pages = {230501},
}

@article{zhangInteractionfreeGhostimagingStructured2019,
	title = {Interaction-free ghost-imaging of structured objects},
	volume = {27},
	issn = {10944087},
	doi = {10.1364/oe.27.002212},
	number = {3},
	journal = {Optics Express},
	publisher = {The Optical Society},
	author = {Zhang, Yingwen and Sit, Alicia and Bouchard, Frédéric and Larocque, Hugo and Grenapin, Florence and Cohen, Eliahu and Elitzur, Avshalom C. and Harden, James L. and Boyd, Robert W. and Karimi, Ebrahim},
	year = {2019},
	pages = {2212},
}

@article{salihProtocolDirectCounterfactual2013,
	title = {Protocol for direct counterfactual quantum communication},
	volume = {110},
	url = {http://arxiv.org/abs/1206.2042},
	doi = {10.1103/PhysRevLett.110.170502},
	number = {17},
	journal = {Phys. Rev. Lett.},
	author = {Salih, Hatim and Li, Zheng-Hong and Al-Amri, M. and Zubairy, M. Suhail},
	year = {2013},
	pages = {170502},
}

@article{dickeInteractionfreeQuantumMeasurements1981,
	title = {Interaction-free quantum measurements: {A} paradox?},
	volume = {49},
	doi = {https://doi.org/10.1119/1.12592},
	journal = {Am. J. Phys.},
	author = {Dicke, Robert H.},
	year = {1981},
	pages = {925--930},
}

@article{elitzurQuantumMechanicalInteractionfree1993,
	title = {Quantum mechanical interaction-free measurements},
	volume = {23},
	doi = {https://doi.org/10.1007/BF00736012},
	journal = {Found Phys},
	author = {Elitzur, Avshalom and Vaidman, Lev},
	year = {1993},
	pages = {987--997},
}

@article{whiteInteractionFreeImaging1998,
	title = {"{Interaction}-{Free}" {Imaging}},
	volume = {58},
	doi = {10.1103/PhysRevA.58.605},
	number = {1},
	journal = {Phys. Rev. A},
	author = {White, Andrew G and Mitchell, Jay R and Nairz, Olaf and Kwiat, Paul G},
	year = {1998},
	pages = {605--613},
}

@article{lemosQuantumImagingUndetected2014,
	title = {Quantum imaging with undetected photons},
	volume = {512},
	issn = {14764687},
	doi = {10.1038/nature13586},
	number = {7515},
	journal = {Nature},
	publisher = {Nature Publishing Group},
	author = {Lemos, Gabriela Barreto and Borish, Victoria and Cole, Garrett D. and Ramelow, Sven and Lapkiewicz, Radek and Zeilinger, Anton},
	year = {2014},
	pages = {409--412},
}

\section*{Appendix}
\appendix
\section{Methods}\label{sec:methods}
\subsection{Experimental setup}

The quantum interrogation tasks explored in this work were implemented on a fully programmable multimode integrated interferometer, or UPP. We make use of the cloud-accessible Ascella Quantum Processing Unit (QPU) provided by the Quandela Cloud service \cite{maringVersatileSinglephotonbasedQuantum2024}. This QPU consists of a 12-mode reconfigurable universal interferometer \cite{reckExperimentalRealizationAny1994,clementsOptimalDesignUniversal2016}, equipped with an on-demand quantum dot single-photon source and superconducting nanowire single-photon detectors at each output. Specifically, a quantum dot source produces single photons at 928 nm, which are injected into the 12-mode interferometric circuit. The chip’s optical circuit is a universal linear-optical network implemented with a mesh of evanescent-field couplers and thermo-optic phase shifters. In total, 132 directional couplers and 126 phase modulators are integrated in a rectangular mesh architecture, which can approximate any $12\times12$ unitary transformation on the creation operators associated with input modes \cite{maringVersatileSinglephotonbasedQuantum2024}, using the optimal decomposition proposed by Clements et al. \cite{clementsOptimalDesignUniversal2016}. These integrated beam splitters are nominally balanced (50/50 splitting), and the phase shifters allow arbitrary relative phases to be programmed, so that by appropriate setting of these elements one can realize any linear interferometer configuration on up to 12 modes. In practice, small fabrication imperfections lead to errors in the implemented unitaries \cite{maringVersatileSinglephotonbasedQuantum2024}. To mitigate these errors, a software transpilation process uses a machine-learning optimization to compile target circuits into precise phase settings \cite{fyrillasScalableMachineLearningassisted2024}. This calibration ensures high-fidelity implementation of the intended interferometer on the chip. The reconfigurability and stability of the monolithic device obviate the need for active path-length stabilization that would be required in a long interferometric setup on an optical table.\par
We interfaced with the photonic processor through the Perceval software framework \cite{heurtelPercevalSoftwarePlatform2023}. Perceval allows users to design linear optical circuits and send them as programs to the QPU. Input states, such as a single photon in a given mode, are specified in the software, and the platform executes the corresponding configuration on the chip remotely. For each programmed circuit, the QPU returns detection events from its 12 output channels, up to a user-specified number of trials (shots). In our experiments, we typically requested on the order of $10^6$ single-photon trials for each configuration to accumulate statistics. In parallel, Perceval can also classically simulate the expected outcomes of the photonic circuit. We performed such simulations using the Strong Linear Optical Simulation (SLOS) backend \cite{heurtelPercevalSoftwarePlatform2023}, which efficiently computes the expected photon number distribution given the interferometer’s unitary matrix. We used these simulations in App.\ref{appendix:robustness} to investigate the theoretical robustness of multimode IFM schemes to errors in individual optical components. \par
The presence of absorptive classical objects within the interferometer was emulated following the approach used in Ref.\cite{giordaniExperimentalCertificationContextuality2023}. An “object” obstructing an interferometer arm is modelled by diverting that path to a dedicated single-photon detector that heralds the photon’s absorption. In the physical chip, this is achieved by programming the interferometer such that the mode corresponding to the object’s position is routed entirely to an output port connected to a detector, and not allowed to interfere with other paths. A click in that detector heralds absorption by the object. All such configurations are implemented on the same photonic chip, which provides a stable and uniform platform for all experiments. We note that in a free-space optical setup, physically placing or removing objects and maintaining alignment for multiple trials would be challenging, whereas here the reconfigurability of the chip enables rapid, reliable toggling of object presence in different modes.\par 
Using the above methods, we implemented various IFM circuits on the Ascella chip and measured the frequencies of different detection outcomes. From the raw detection counts, we estimate the efficiency $\eta$ of each IFM scheme as the fraction of successful interaction-free detections out of all trials that resulted in a detection or absorption. For a single-object IFM, this corresponds to the standard definition $\eta = P_{\rm IFM}/(P_{\rm IFM}+P_{\rm abs})$, detailed in Sec.\ref{section:EV}. We applied a  generalised efficiency for multi-object scenarios, introduced later in Sec.\ref{section:our_multiple_object}. The statistical uncertainty in $\eta$ was estimated following an error mitigation technique detailed in App.\ref{appendix:errorbars}. In comparing experiment to theory, the primary sources of deviation are imperfections in the implemented unitary - e.g. slight miscalibration of phase shifts or splitting ratios - and detector dark counts. Each circuit was compiled and executed using the same hardware with minimal adjustments, which ensures a fair comparison between
different IFM schemes under consistent conditions.

\subsection{Error mitigation}\label{appendix:errorbars}

\begin{figure}
    \centering
    \includegraphics[width=.95\linewidth]{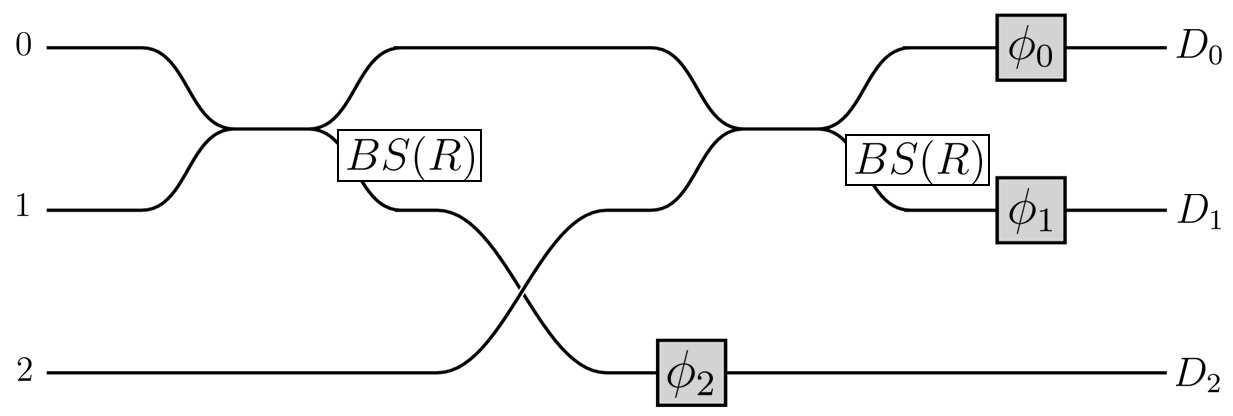}
    \caption{Error-mitigated circuit implemented to estimate the efficiency of the standard EV task. Phase shifters with random phases $\phi_0$, $\phi_1$ and $\phi_2$ are added to each mode just before detection. For each reflectivity $R$ value, $M=40$ circuits were sampled from, each with a different set of phases randomly sampled from a uniform distribution $\phi_i \in [0,2\pi[$. These additional phases do not alter the output probabilities of the EV circuit, but help the compilation procedure in Perceval converge to a higher fidelity unitary.}
    \label{fig:mitigation}
\end{figure}

\begin{figure*}
    \centering
    \subfigure[]{
    \includegraphics[width=.475\linewidth]{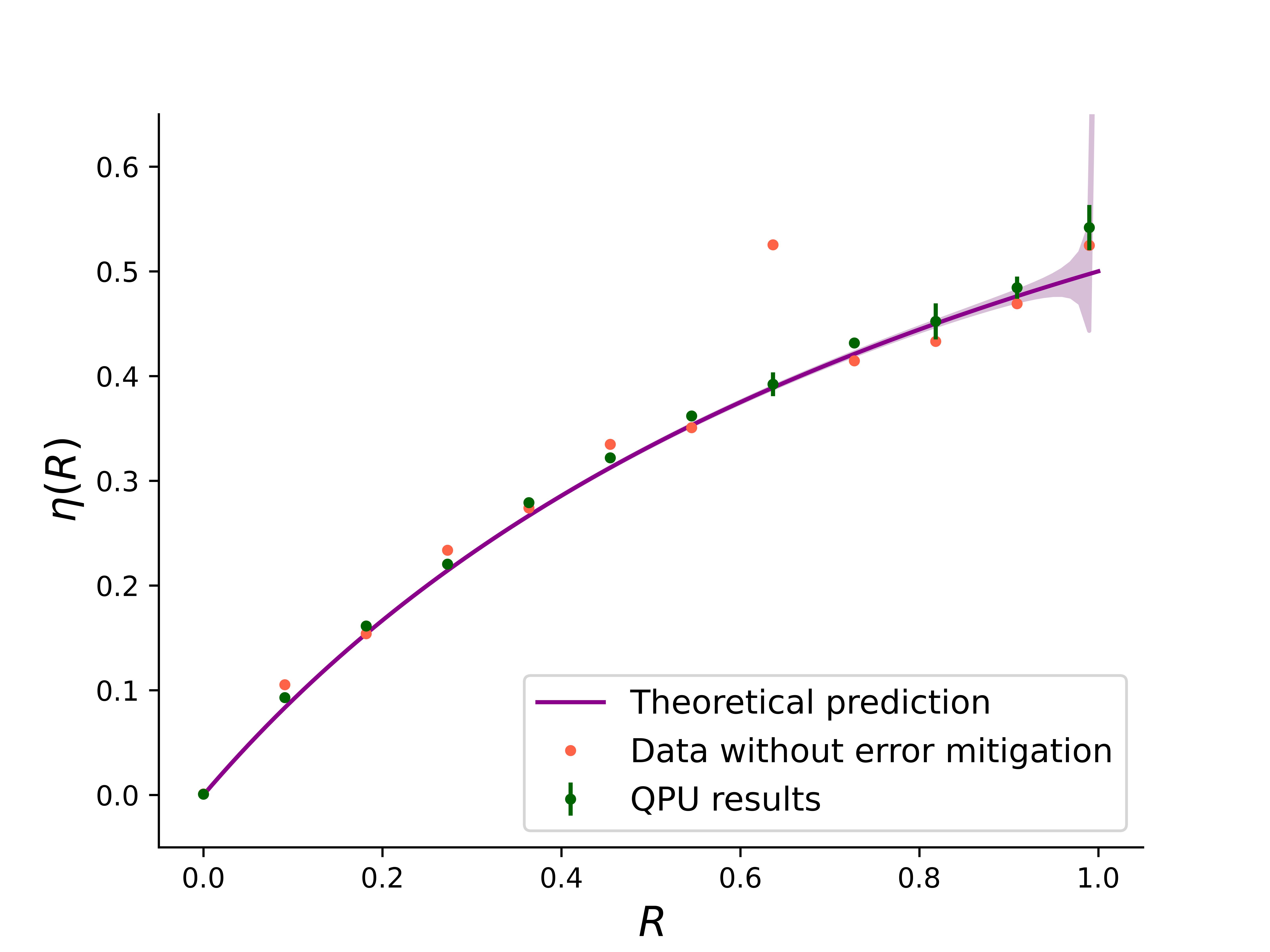}\label{EV_data}}
    \subfigure[]{
    \includegraphics[width=.475\linewidth]{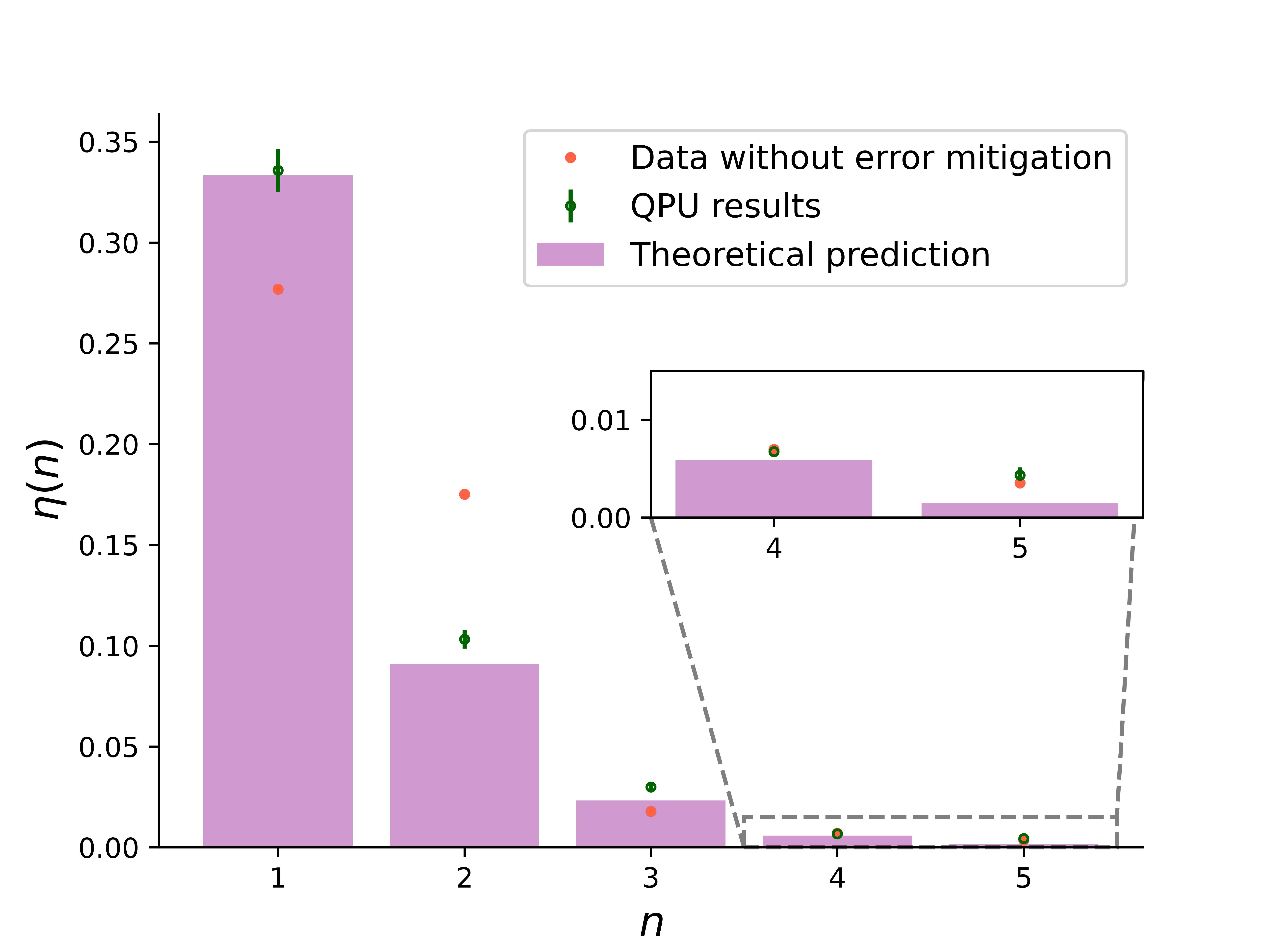}\label{multibomb_data}}
    \caption{Experimental data collected for (a) the EV single-object IFM task and (b) the multi-object IFM task. In (a), we represent the protocol's efficiency $\eta(R)$ as function of the beam-splitter reflectivity $R$; in (b), the efficiency $\eta(n)$ as a function of the number $n$ of absorbing objects. In both graphs, the error mitigated data shown in green correspond to the estimation of $\eta$ obtained by averaging photon counting statistics over $40$ random circuits. For comparison, we represent in red an estimation of $\eta$ where no error mitigation is performed (see details in main text).}
    \label{fig:data}
\end{figure*}

For all IFM schemes implemented in this work, $\eta$ was estimated from order $N\approx10^6$ photon count samples. For these collected statistics, the error due to poissonian noise in single-photon counts was found to be insignificant compared to other sources of noise for all schemes. A second source of error are detector dark counts. Due to the nature of cloud-based services like Quandela cloud, direct control of the experimental apparatus is not possible, so it was not possible to characterize this source of noise.\par
We found that an important source of noise was related to optical component imperfections and the programming of the QPU. The physical device consists of a $12\times12$ mesh of 126 voltage-controlled thermo-optic phase shifters and 132 directional couplers \cite{maringVersatileSinglephotonbasedQuantum2024}, arranged in Mach-Zehnder units in the Clements rectangular architecture \cite{clementsOptimalDesignUniversal2016}. The chip exhibits various imperfections. For instance, while in principle the directional couplers are fabricated to behave as perfectly balanced beamsplitters, in practice, the average reflectivity observed is of $56.7\%$ \cite{maringVersatileSinglephotonbasedQuantum2024}. Furthermore, differences in the optical coupling to the chip or in detection efficiencies lead to inhomogeneous input and output optical transmissions across the different ports. Finally, passive phases due to fabrication defects and crosstalk induced by reconfigurable components may affect the effective phase induced by the tunable phase shifters. These limitations are mitigated in Perceval using an iterative machine-learning procedure \cite{fyrillasScalableMachineLearningassisted2024}. A global optimization step compiles user-provided photonic circuits into phase shift values in the interferometer, simultaneously adjusting all phase shift values, and a transpilation step converts these values into voltages to apply on the thermo-optic phase shifters. This calibration procedure highly compensates for physical imperfections, improving unitary fidelity. \par
For some of the implemented circuits, we observed a systematic error in the estimated value of $\eta$. Furthermore, we found that this bias was contingent on the calibration of the chip performed on-site - from one calibration routine to the next, different implemented schemes exhibited different biases. We concluded that these biases were related to compilation and/or transpilation artefacts - for some circuits, the machine-learning iteration was converging to a set of phase shifter values that resulted in a lower fidelity unitary. Therefore, on top of Perceval's built-in calibration procedure, we applied an error mitigation procedure to further reduce this systematic bias. For each IFM scheme, we provided $M = 40$ circuits to Perceval. In each circuit, we added an additional, random phase picked from a uniform distribution $[0,2\pi[$ in all modes, just before detection - see Fig.\ref{fig:mitigation} for an example with the standard EV scheme. These additional phases do not affect the probability distribution of photons at the output of the device, but they randomize the initial parameters input to the compilation procedure. This helps the algorithm converge to a higher fidelity unitary.
We extracted $N\approx10^6$ samples from each of the $40$ circuits, with which we obtained one estimate of $P_{\text{IFM}}$ and $P_{\text{{abs}}}$ per circuit. We then performed an ensemble average over the $M=40$ estimates of each probability to get an estimation of their mean, with the standard deviation of the mean being $\sigma_M = \frac{\sigma}{\sqrt{M}}$. The final value for $\eta$ was then computed using these mean estimations, with the error bars obtained from standard error propagation. In Fig.\ref{fig:data}, we compare the final, error mitigated results for both IFM schemes, reported in green, with the estimation of $\eta$ obtained without error mitigation, in red. Notably, we can see deviations as large as $\approx40\%$ relative error from the theoretical value for some of the raw data points, which are greatly improved with the mitigation procedure.\par

\subsection{Dark and object detector counts in unobstructed interferometers}\label{appendix:data_without_object}

\begin{figure*}[htbp]
    \centering
    
    \subfigure[]{
        \includegraphics[width=0.4\textwidth]{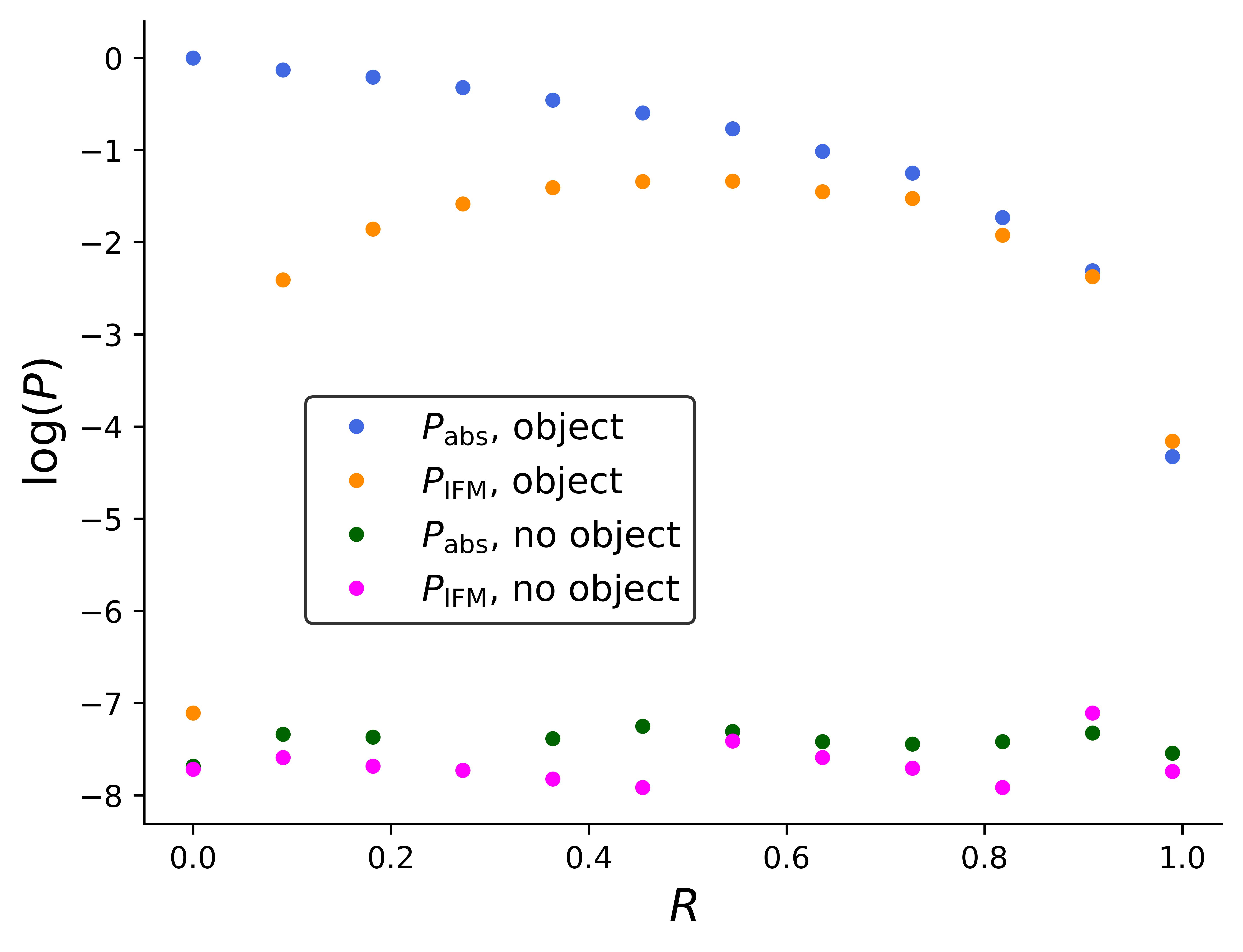}
    \label{fig:no_obj_EV}
    }
    \hspace{0.02\textwidth}
    \subfigure[]{
        \includegraphics[width=0.4\textwidth]{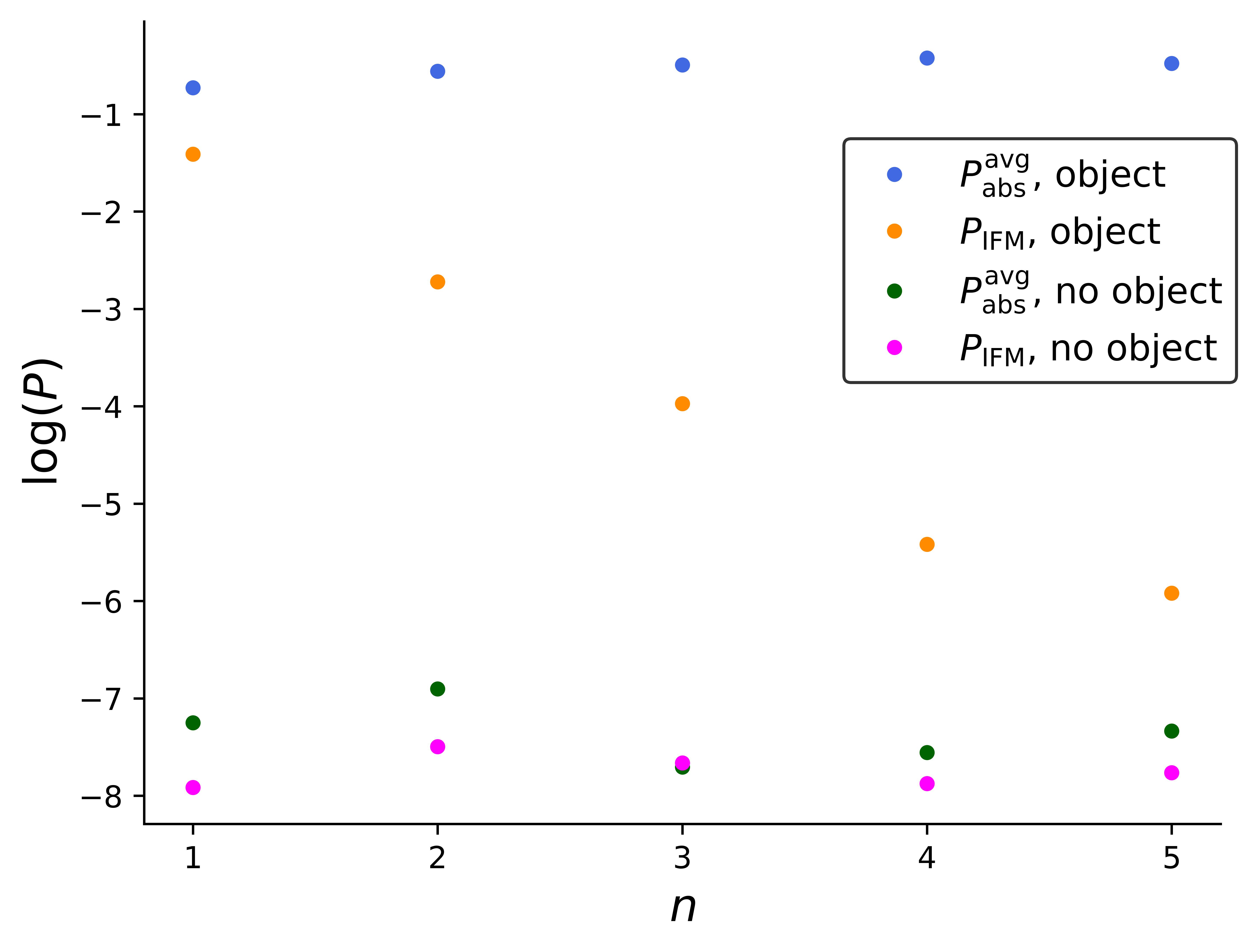}
    \label{fig:no_obj_multi}
    }
    \caption{Data for IFM circuits where the absorbing object is removed. (a) Photon counting probabilities, in logarithmic scale, for the dark ($P_{\text{IFM}}$) and object detectors ($P_{\text{abs}}$) of the standard EV scheme represented in Fig.\ref{fig:EV_scheme}, in the scenarios where the mode permutation emulating the object is either present or removed, as a function of the variable reflectivity $R$. (b) Similar data for the schemes with $n$ objects, such as the one in Fig.\ref{fig:two_bomb_circuit} for the case of $n=2$ objects. $P_{\text{abs}}^{\text{avg}}$ corresponds to the average photon counting probability over all object detectors. Statistical error bars are not visible in both graphs.}
    \label{fig:no_obj}
\end{figure*}

In order to confirm that the observed dark counts in the different IFM setups were due to the presence of the object(s) in the interferometers and not simply due to noise, we performed identical experiments to the ones reported in the main text but without introducing the absorber. Take, for example, the interferometer in Fig.\ref{fig:EV_scheme}. In ideal conditions, in the absence of the mode permutation emulating the absorbing object, all input photons should exit at the light port detector $D_0$, with destructive interference occurring at the dark port detector $D_1$, and with no photons being directed to the object detector $D_2$. Thus, we should find a null absorption probability at $D_2$, $P_{\text{abs}}=0$, and no IFM counts at $D_1$, $P_{\text{IFM}}=0$. A similar reasoning can be applied to the schemes for multiple objects, such as the one in Fig.\ref{fig:two_bomb_circuit}. In practice, experimental errors such as imperfect optical components, below-unity fidelity of the implemented unitaries on the chip or detector dark counts lead to leakages towards the dark detector, as well as the object detector(s), even in the absence of the mode permutation(s) redirecting to the object detector(s). Nevertheless, a successful experimental demonstration of an IFM should achieve a much larger number of photon counting events at the dark and object detector(s) in the presence of obstruction than in its absence.

In Fig.\ref{fig:no_obj_EV}, we compare the measured photon counting probabilities $P_{\text{IFM}}$ and $P_{\text{abs}}$, at $D_1$ and $D_2$, respectively, obtained for the EV interferometer in Fig.\ref{fig:EV_scheme} in the scenarios with and without the object, that is, with and without the mode permutation between modes 1 and 2. We display the data in logarithmic scale for a clear comparison of their orders of magnitude. We can conclude that the dark and object counts are of a significantly higher order of magnitude in the scenario where the object is present, except, arguably, for the point closest to $R=0$, where the dark counts are comparable in both scenarios. Note, however, that this behaviour is expected, since, as we approach the limit $R \to 0$, the probability of dark counts is vanishingly small, as most input photons are absorbed by the object. We can therefore conclude that the observed counts in the presence of the object in fact demonstrate a successful IFM. Fig.\ref{fig:no_obj_multi} shows similar results for the setups with $n=1,...,5$ objects. When considering schemes with 2 or more objects, there are several configurations of present and absent objects we can consider. For example, in the 5 object scheme, we can consider that all objects are removed, or only the lowermost one, or only the third one, and so on. However, note that, as soon as the $i^{th}$ object is removed, all objects below it in the scheme are rendered undetectable, making the scheme equivalent to the one with only $i-1$ objects. Therefore, for the sake of simplicity, we consider here only the configurations where either all objects are present, or all are removed. $P_{\text{IFM}}$  corresponds to the photon count probability at the detector which simultaneously heralds the IFM of all objects in a single click. For the scenario with $n$ objects present, we consider the average probability of object absorption events over all object detectors $P_{\text{abs}}^{\text{avg}} = (\sum_i^nP_{\text{abs}}^i)/n$. The dark count and average object count probabilities in the absence of all objects across the five different schemes are at least one order of magnitude smaller than the corresponding probabilities measured with the object(s) present. These results validate the successful demonstration of IFMs of up to five objects in the Ascella photonic processor.

\section{Noise robustness in multimode single object interrogation}\label{appendix:robustness}

\begin{figure}
    \centering
    \includegraphics[width=\linewidth]{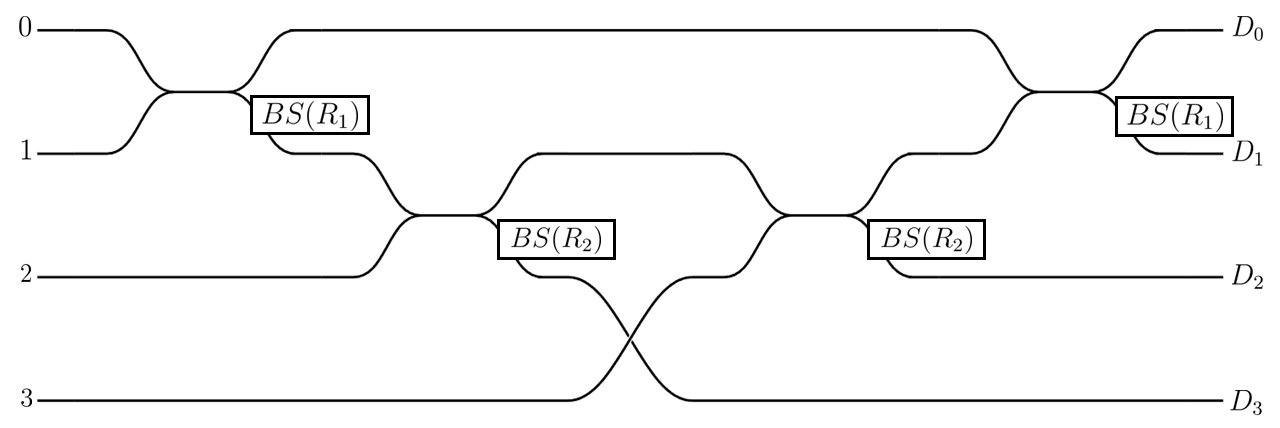}
    \caption{Circuit simulated in Perceval for a $m=3$ mode setup with two tunable reflectivities $R_1$ and $R_2$, used to assess the performance of multimode IFM setups under errors in optical components. For general $m$, we use $m+1$ modes, placing cascading pairs of BSs in the first $m$ modes and using the last mode to emulate the presence of an absorptive object.}
    \label{fig:m3_circuit}
\end{figure}
\begin{figure}[h]
    \centering
    \includegraphics[width=.45\textwidth]{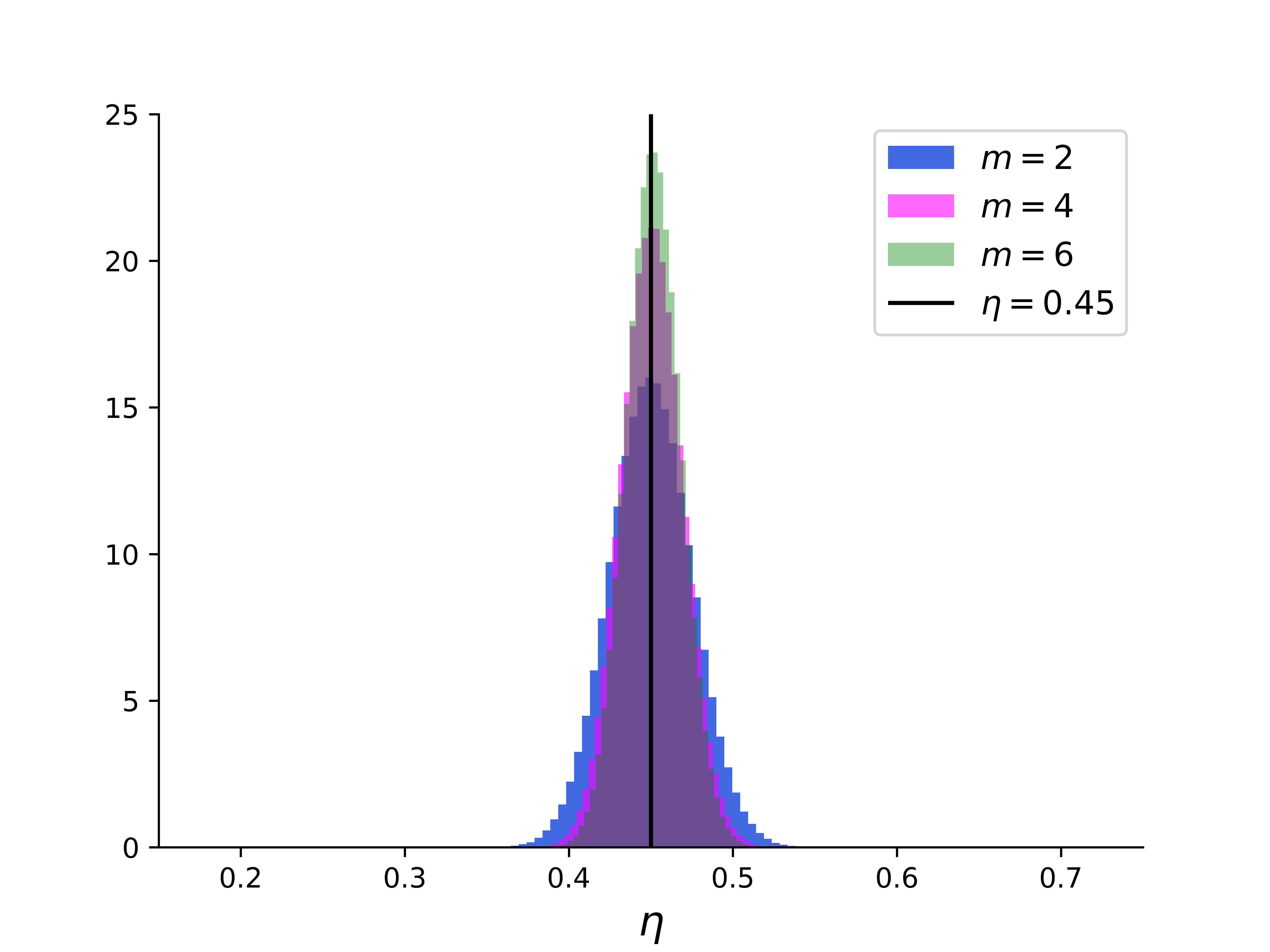}
    \includegraphics[width=.45\textwidth]{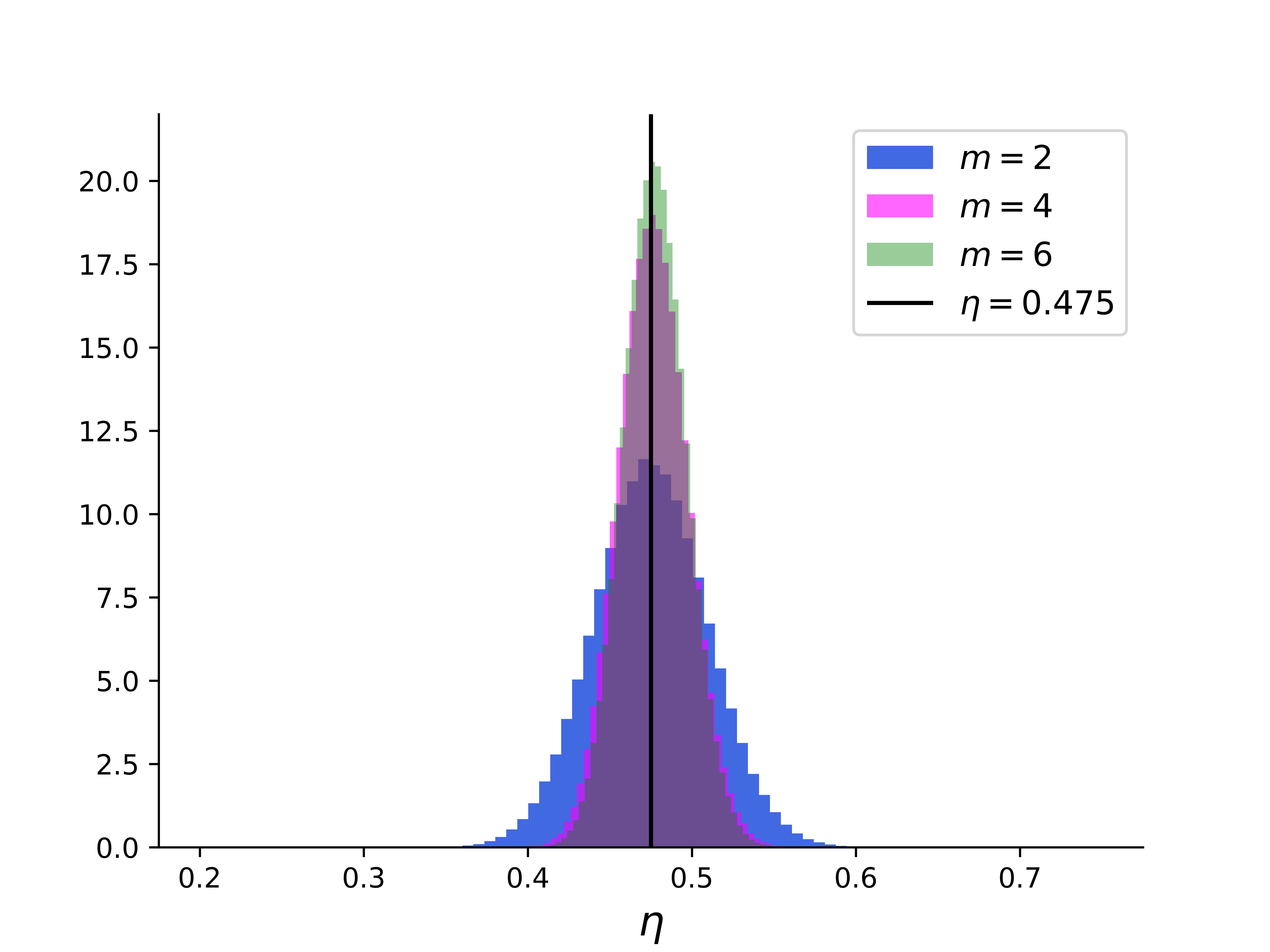}
    \includegraphics[width=.45\textwidth]{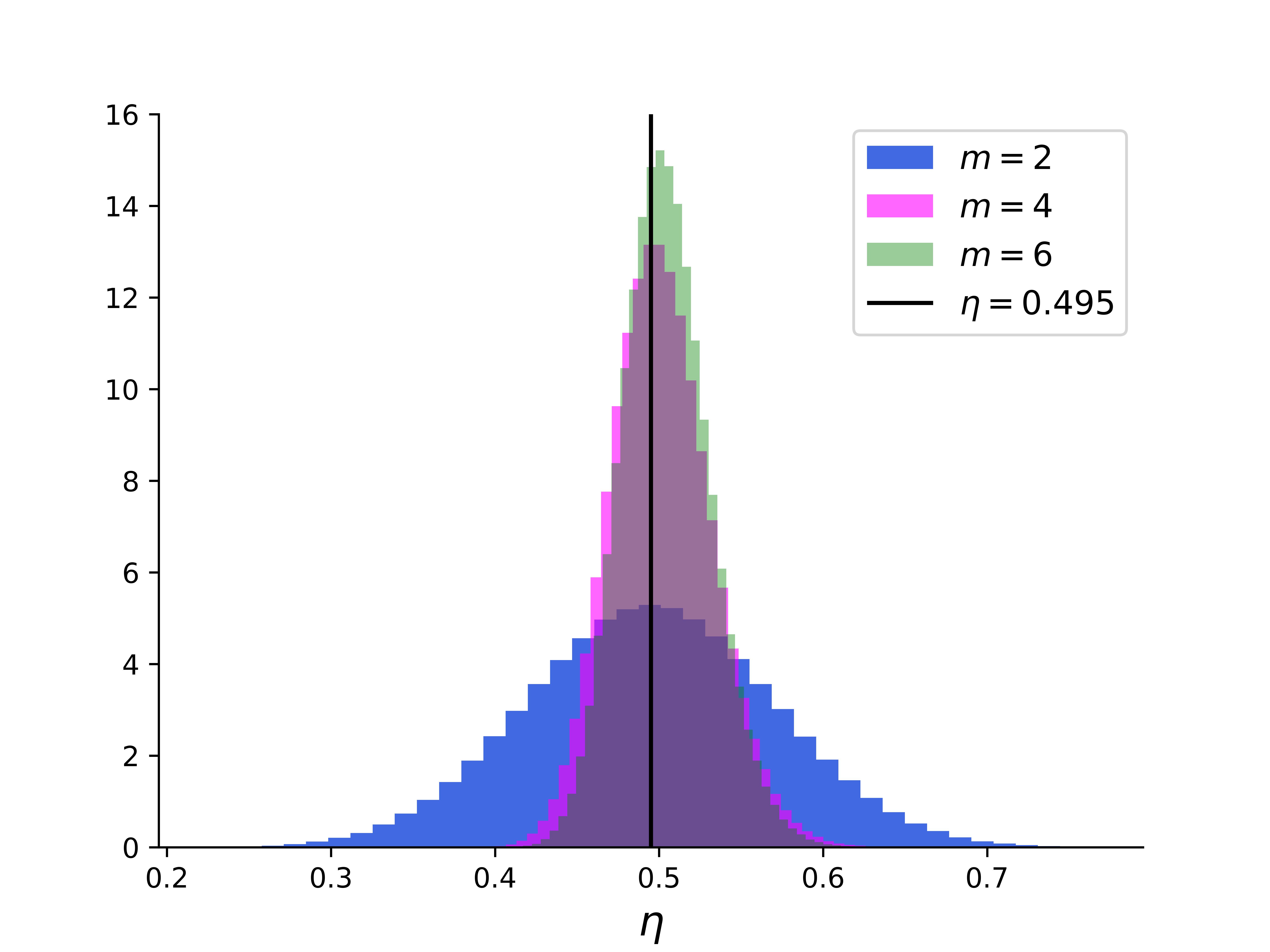}
    \caption{Histograms of simulated efficiency data of $m$-mode IFM setups, such as the one in Fig.\ref{fig:m3_circuit} for $m=3$, under Gaussian errors in the reflectivity of each beamsplitter.}
    \label{fig:hists}
\end{figure}
\begin{figure}
    \centering
    \includegraphics[width=\linewidth]{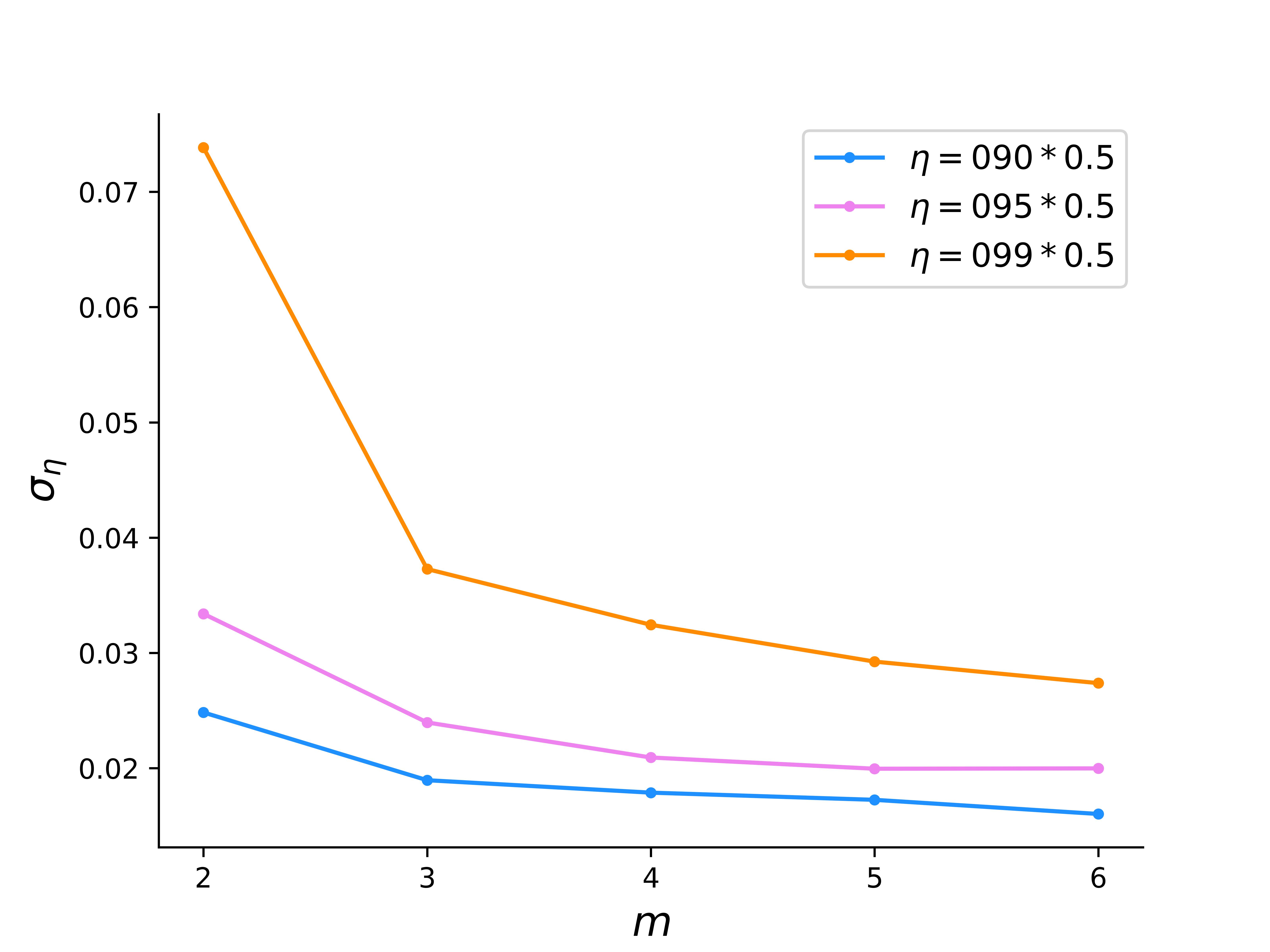}
    \caption{Standard deviation of efficiency of multimode setups, as a function of the number of modes $m$, computed from the histograms in Fig.\ref{fig:hists} for different target efficiencies close to the $0.5$ ceiling.}
    \label{fig:std_eta}
\end{figure}
We have seen in Sec.\ref{section:EV} how, in the standard EV protocol, trying to approach the $50\%$ efficiency limit means pushing the BS reflectivity $R$ very close to 1, which, as discussed, makes the setup more sensitive to slight mismatches in the interferometer. In light of this, a multimode approach could offer robustness advantages, since in an $m$-path scheme, one can achieve a high effective reflectivity through the interferometer design using individual BSs each with a smaller reflectivity. For instance, consider the 4-mode circuit in Fig.\ref{fig:m3_circuit}, with $U_3 = (BS(R_1)\otimes I)(I \otimes BS(R_2))$ (i.e., the first beamsplitter acts only on the top two modes, and the second one on the bottom two modes), and an additional mode to emulate the object's presence. According to eq. \eqref{eq:eta_U}, the efficiency is given by
\begin{equation}
    \eta  =\frac{1 - \tilde{T}}{2 - \tilde{T}}
\end{equation}
\noindent with $\tilde{T} = T_1T_2$. This is equivalent to the standard EV setup, with an effective reflectivity $\tilde{R} = R_1 + R_2 - R_1R_2$, and it can be shown that $\eta$ is more sensitive to reflectivity errors when both $R_1$ and $R_2$ approach 1. However, with this setup, a given target $\tilde{R}$, and thus, a given target $\eta$, can be achieved with individually smaller $R_1$ and $R_2$, when compared to the simple EV scheme with a single tunable $R$ parameter. This could result in a less significant error arising from reflectivity mismatches between $U_m$ and $U_m^{\dagger}$, leading to smaller fluctuations in the efficiency close to the optimal value. Note that this is only true if we can assume that the errors in $R_1$ and $R_2$ are random and uncorrelated.\par
We checked this intuition with a numerical noise model. We locally simulated an IFM for $m=2,3,4,5,6$ modes using the SLOS backend in Perceval (see Sec.\ref{sec:methods}), generalizing the circuit in Fig.\ref{fig:m3_circuit} for each $m$, using $m$ modes with $m-1$ pairs of beamsplitters and an additional, lowermost mode to emulate absorption. The simulations were performed for target values of $\eta_{\text{target}} = 0.9*\eta_{\text{limit}}$, $\eta_{\text{target}} = 0.95*\eta_{\text{limit}}$ and $\eta_{\text{target}} = 0.99*\eta_{\text{limit}}$, with $\eta_{\text{limit}} = 0.5$ the efficiency ceiling. For each $\eta_{\text{target}}$, we chose a corresponding set of target reflectivities for each BS (for simplicity, we set all reflectivities to be equal), and allowed small Gaussian fluctuations around the target value (standard deviation of 0.03). Figure \ref{fig:hists} shows the results for $10^6$ estimations of the efficiency under this noise model for each $\eta_{\text{target}}$, exemplified for $m=2,4,6$. We found that, for the two-mode case, the variance in the achieved $\eta$ grows significantly as the target approaches $\eta_{\text{limit}}$, implying that the slightest imbalance can spoil the interference, and corroborating the assumption that reflectivity mismatches are responsible for the deviations observed in Fig.\ref{fig:eta(R)_plot}. In contrast, for higher $m$, the efficiency estimates remained much more tightly distributed around the target values even in the presence of comparable perturbations. In other words, the larger interferometers yielded smaller fluctuations in $\eta$ near the optimal regime, suggesting that distributing the interference across multiple modes can make the protocol less susceptible to single BS imperfections.\par
This trend can be seen more clearly in Fig.\ref{fig:std_eta}, where we plot the standard deviation of the histograms of $\eta$ for each target value, as a function $m$. The advantage of using more beamsplitters of smaller reflectivity each is countered by the additional noise due to the introduction of more components, so that the decrease in the fluctuations of $\eta$ is not so significant for higher $m$. \par 
Note that the assumption that errors in the individual BSs are independent does not hold for the UPP; we cannot perfectly isolate the optical components out of the $12 \times 12$ grid that correspond to each BS, and the values of individual phase shifters are all simultaneously tweaked in a global, iterative optimization, resulting in correlated errors. This physical platform is thus not suited to test the validity of our noise model. A bulk-optics setup, for example, would be more appropriate to evaluate the robustness of setups with higher number of modes.\par

\section{Optimal efficiency with multiple objects}\label{appendix:simulations}

\begin{figure}[h]
    \centering
    \includegraphics[width=\linewidth]{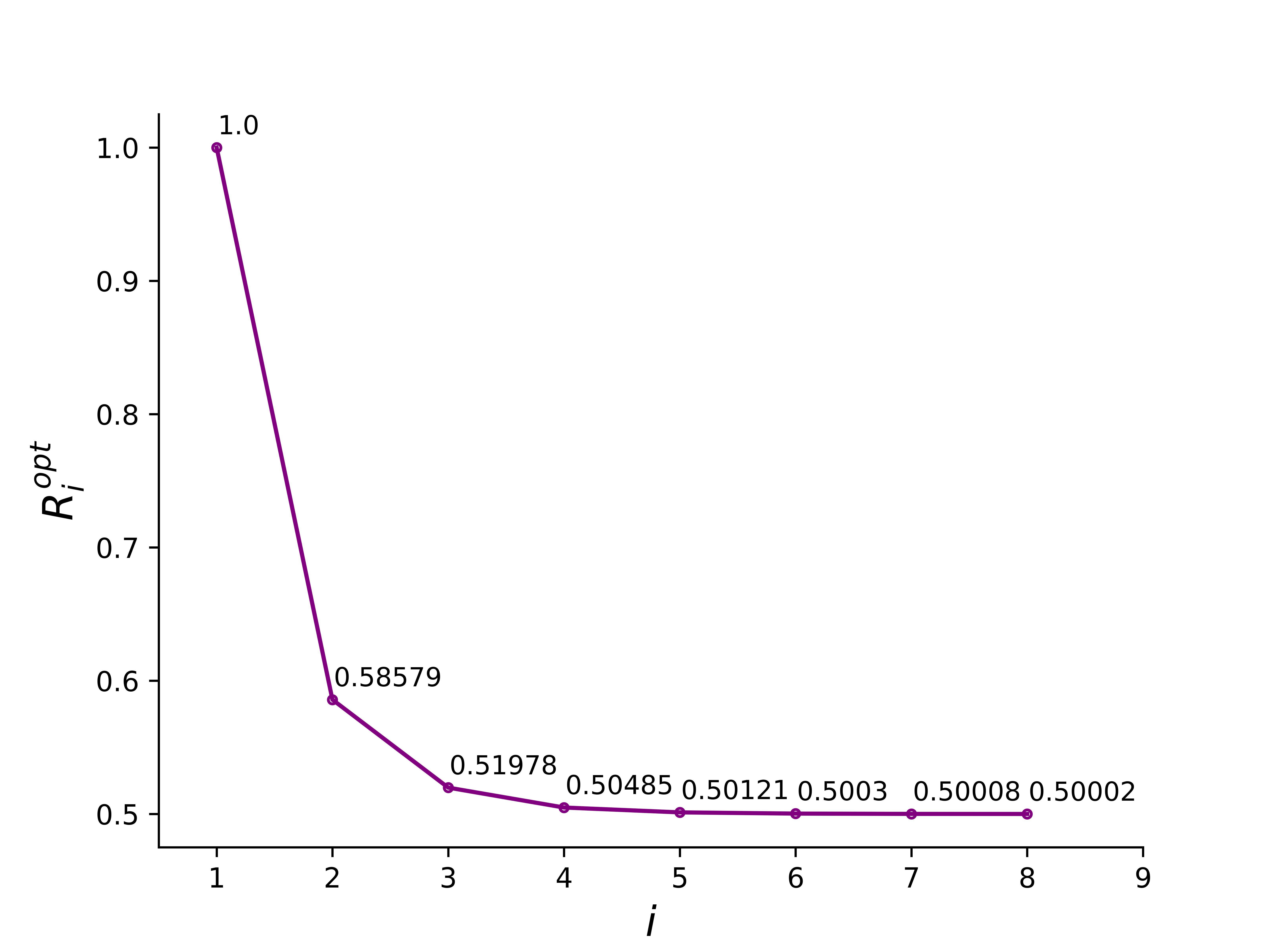}
    \caption{Numerically determined values of $R_{i}^{\text{opt}}$, the reflectivity of each pair of beamsplitters of the $i^{th}$ EV box which maximize $\eta$ for the multiple object, interaction-free measurement scheme. See details in main text.}
    \label{fig:optimal_R}
\end{figure}
As discussed in Sec.\ref{section:multiple_object_UPP}, the experimental results from the QPU for the $n$-object setup in Fig.\ref{fig:n_objects} were obtained by fixing the reflectivity $R_i$ of the pair of beamsplitters in the $i^{th}$ EV box to be all equal to $R_i=0.5$. However, this choice of $R_i$ does not yield the optimal efficiency attainable for the $n$ object IFM scheme. For example, in the $n=1$ scheme, corresponding to the standard EV task, we discussed in Sec.\ref{section:EV} how the optimal value for $\eta$ of $0.5$ is asymptotically approached as $R \to 1$, with the caveat that $R$ cannot be exactly equal to unity, otherwise the possibility of interaction with the object is null, and an IFM cannot be performed. We further discussed that, as $R\to 1$, the probability $P_{\text{IFM}}$ of a click at the dark detector vanishes, with most probes ending up at the light detector instead. In this situation, an increasing number of trials need to be performed on average before an IFM outcome is observed.\par
The optimal $R_i^{\text{opt}}$ values yielding maximum efficiency were numerically determined for the IFM schemes with $n = 2$-$8$. Curiously, we found that the optimal choice $R_i^{\text{opt}}$ for the beamsplitters in the $i^{th}$ EV box is the same regardless of the number of objects we consider. For instance, the optimal choice for $R_1^{\text{opt}}$, corresponding to the EV box in the two uppermost modes, is fixed at $R_1\to 1$, whether we consider it in the $n=1$ scheme, the $n=2$ scheme, the $n=3$ one, and so on. All other optimal $R_i^{\text{opt}}$ values are likewise independent of the scheme. These are represented in Fig.\ref{fig:optimal_R}. We see that, to maximize $\eta(n)$, the reflectivity of the beamsplitters of each EV box, from the top to the bottom mode, seems to asymptotically approach $R=0.5$.\par
In Fig.\ref{fig:eta(n)_plot}, we represented the optimal efficiency for the $n=1,..,5$ IFM schemes computed using the $R_i^{\text{opt}}$ from Fig.\ref{fig:optimal_R}. However, we did not use this set of $R_i$ when acquiring experimental data from the QPU, and instead opted to set $R_i=0.5$ for all boxes. The principal reason for this choice has to do with the vanishingly small probability $P_{\rm IFM}(n)$ of the counterfactual outcome for the $R_i^{\text{opt}}$ configuration, which makes it difficult to accumulate photon counting statistics at the IFM detector using the optimal setup. If we were to program the QPU using $R_1$ close to unity, many trials would be required in order to obtain a statistically significant sample of all possible outcomes for the circuits with larger $n$. For example, setting $R_1\approx 0.99$ and $R_i = R_i^{\text{opt}}$ for $i = 2,..,5$ in the $n=5$ scheme, the $P_{\text{IFM}}$ probability, corresponding to obtaining a click at the dark detector $D_5$ in a given trial, would be only of the order of $10^{-5}$. Gathering enough samples to properly estimate such a small probability was not feasible when using the cloud-accessible QPU. This motivated our choice of $R_i=0.5$, which corresponds to the configuration that maximizes $P_{\rm IFM}(n)$. Not only did it simplify the configuration of the QPU, it facilitated the collection of meaningful statistics for the larger $n$ schemes. This illustrates the practical considerations discussed in Sec.\ref{section:our_multiple_object}, where we argued that it is not clear whether maximizing the efficiency is always desirable in practical settings, since it comes at the price of requiring an exponentially increasing amount of resources to obtain a reliable IFM detection when increasing the number of objects. Furthermore, as can be seen from Fig.\ref{fig:eta(n)_plot}, the difference between the optimal expected efficiency and the one corresponding to the symmetric $R_i=0.5$ configuration is smaller as more objects are added, and the collected data still capture the meaningful behaviour of quickly decreasing efficiency with increasing $n$.

\end{document}